\documentclass[sigconf,10pt,nonacm]{acmart}
\AtBeginDocument{%
  }

\usepackage{pgfplots}
\usepackage{graphicx}
\usepackage{caption}
\usepackage{subcaption}
\usepackage{booktabs}
\usepackage[ruled,linesnumbered,noend]{algorithm2e}
\usepackage{amsmath}
\usepackage{hyperref}
\usepackage{enumitem}
\usepackage{titlesec}
\usepackage{multirow}
\usepackage{multicol}
\usepackage{xcolor}

\definecolor{myred}{RGB}{172,0,0}
\definecolor{myblue}{RGB}{0,112,192}
\definecolor{mygreen}{RGB}{78,167,46}
\definecolor{myorange}{RGB}{233,113,50}
\titlespacing*{\section}
  {0pt}{0.5em}{0.5em}  
\titlespacing*{\subsection}
  {0pt}{0.5em}{0.5em}  
\titlespacing*{\subsubsection}
  {0pt}{0.5em}{0.5em}

\begin{document}

\title{MobileRAG: A Fast, Memory-Efficient, and Energy-Efficient Method for On-Device RAG}


\author{Taehwan Park}
\orcid{0009-0009-4160-1092}
\affiliation{%
  \institution{Korea Advanced Institute of Science and Technology}
  \city{Daejeon}
  \country{Republic of Korea}
}
\email{always_hwan@kaist.ac.kr}

\author{Geonho Lee}
\orcid{0009-0008-7133-491X}
\affiliation{%
  \institution{Korea Advanced Institute of Science and Technology}
  \city{Daejeon}
  \country{Republic of Korea}
}
\email{ghlee5084@kaist.ac.kr}

\author{Min-Soo Kim}
\authornote{Corresponding author.}
\orcid{0000-0002-7852-0853}
\affiliation{%
  \institution{Korea Advanced Institute of Science and Technology}
  \city{Daejeon}
  \country{Republic of Korea}
}
\email{minsoo.k@kaist.ac.kr}


\begin{abstract}
Retrieval-Augmented Generation (RAG) has proven effective on server infrastructures, but its application on mobile devices is still underexplored due to limited memory and power resources. Existing vector search and RAG solutions largely assume abundant computation resources, making them impractical for on-device scenarios. In this paper, we propose MobileRAG, a fully on-device pipeline that overcomes these limitations by combining a mobile-friendly vector search algorithm, \textit{EcoVector}, with a lightweight \textit{Selective Content Reduction} (SCR) method. By partitioning and partially loading index data, EcoVector drastically reduces both memory footprint and CPU usage, while the SCR method filters out irrelevant text to diminish Language Model (LM) input size without degrading accuracy. Extensive experiments demonstrated that MobileRAG significantly outperforms conventional vector search and RAG methods in terms of latency, memory usage, and power consumption, while maintaining accuracy and enabling offline operation to safeguard privacy in resource-constrained environments.

\end{abstract}



\maketitle

\section{Introduction}

Modern smartphones have evolved beyond simple communication tools and now store large volumes of personal data, such as photos, documents, and videos. Users increasingly expect immediate and context-rich retrieval and analysis of this data, yet conventional keyword-based search often struggles with ambiguous or natural-language queries. RAG systems overcome these limitations by coupling similarity-based document retrieval with a LM capable of summarizing or interpreting the retrieved content \cite{NEURIPS2020_6b493230,guo2024lightrag,han2025retrievalaugmentedgenerationgraphsgraphrag}. For instance, queries like \textit{show me pictures from last summer at the beach} or \textit{show me the dessert recipe that I recently downloaded} can be handled more effectively when approximate similarity search methods \cite{10.1145/276698.276876,5432202,8594636} are combined with text-generation techniques. Running these processes on-device is crucial for accessing personal data due to privacy concerns discourage uploading personal photos, documents, and messages to external severs \cite{10.1145/3307334.3326071,10.1145/3626772.3657662}. However, mobile devices face unique constraints: limited memory (typically 4-12GB RAM), battery limitations, and the need to maintain responsiveness for other applications. These constraints require specialized indexing and retrieval techniques that balance accuracy with resource efficiency for large-scale personal data collections.

Existing vector search algorithms, such as IVF \cite{5432202, babenko2014inverted} and HNSW \cite{8594636}, maintain compact indices and support high-speed Approximate Nearest Neighbor Search (ANNS), making them well-suited for data-intensive applications like search engines, recommendation systems, and knowledge discovery tasks. As dataset sizes continue to grow—reaching the billion scale—approaches like DiskANN \cite{jayaram2019diskann} and SPANN \cite{chen2021spann} leverage disk-based storage to accommodate data that exceeds main memory while striving to preserve high recall. Collectively, these methods play a critical role in retrieving relevant items from large-scale, high-dimensional data, ensuring both efficient computation and robust similarity search.

\begin{figure}[t]
\vspace*{-0.2cm}
  \includegraphics[width=\linewidth]{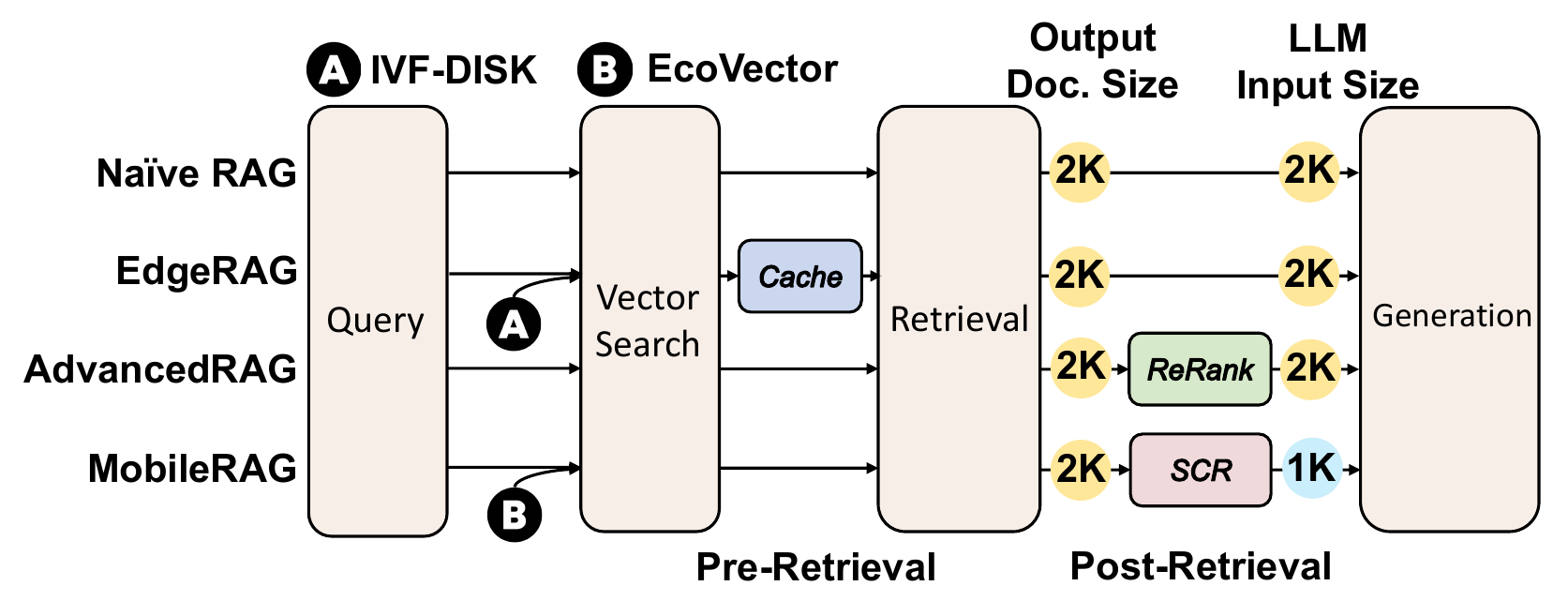}
  \caption{Comparison RAG Methods with MobileRAG.}
  \label{fig:RAG}
  \vspace*{-0.3cm}
\end{figure}

Meanwhile, a range of RAG pipelines employ a two-stage framework where top-$k$ documents are first retrieved for each query, and then passed to the Large Language Model (LLM) for answer generation. In a Naive-RAG \cite{NEURIPS2020_6b493230}, the retrieval outputs are directly fed into the model\,(see Figure~\ref{fig:RAG}); however, more sophisticated variants introduce additional modules for improved accuracy. Re-Ranker-based Advanced RAG \cite{yu2024rankrag,gao2024modular} re-evaluates the initial retrieval results using a secondary model, aiming to refine the candidate set before generation. 
EdgeRAG \cite{seemakhupt2024edgerag} targets pre-retrieval resource constraints using optimized memory management techniques such as IVF-DISK indexing and embedding caching. IVF-DISK partitions embeddings, storing them on disk and loading data on-demand, while caching reduces repeated disk access—lowering both latency and RAM usage. Full-scale LLM deployment on mobile devices is impractical due to limited resources, spurring interest in lighter, parameter-reduced models for on-device use \cite{sanh2019distilbert,jiao2019tinybert,sun2020mobilebert,xin2020deebert}. Many applications now employ Small Language Models (sLM) like MobileBERT or MobileLLM \cite{sun2020mobilebert}, which reduce memory overhead and speed up inference on smartphones \cite{McGowan2021,liu2024mobilellm}. However, these approaches still face three major limitations when deployed on mobile devices.

\textbf{Problem 1 (Memory Footprint Limitations):} Vector search algorithms such as IVF-based methods\,(e.g., IVF, IVF\-PQ, IVF-HNSW) and HNSW typically require the entire index or graph to reside in RAM, leveraging high-performance CPUs for rapid distance computations \cite{5432202,jegou2010product,ge2013optimized,kalantidis2014locally,8594636}. Typically, mobile devices have limited RAM after accounting for the OS, while servers often utilize hundreds of GB. For example, excluding the OS, the Galaxy S24 leaves only 5–6 GB available for applications \cite{McGowan2021}. This limitation can easily trigger out-of-memory issues, making large-scale indices particularly impractical. Furthermore, HNSW-based approaches exacerbate memory issues by requiring the entire graph to remain in memory during searches \cite{8594636}, and even incremental methods face similar challenges. Additionally, integrating Re-Ranker further exacerbates memory constraints, as these components demand additional memory overhead during inference, inflating overall resource usage.

\vspace*{0.1cm}
\textbf{Problem 2 (Power Limitations):} Existing approaches face severe power consumption constraints. Vector search algorithms, particularly IVF-based methods (e.g., IVF, IVFPQ, IVF-HNSW), perform multiple distance computations per candidate within selected clusters, often involving repeated lookups across thousands of vectors in quick succession \cite{jegou2010product,ge2013optimized,kalantidis2014locally}. This significantly increases CPU usage and thus power consumption. Iterative-RAG further compounds this issue by repeatedly invoking an LM, causing sustained high CPU load, rapid overheating, and eventual throttling \cite{yang2024irag}. Once thermal limits are reached, power draw remains elevated, rapidly depleting the battery.

\vspace*{0.1cm}
\textbf{Problem 3 (Latency Limitations):} Standard RAG pipelin\-es suffer from high latency due to both the vector-based retrieval and sLM inference phases \cite{NEURIPS2020_6b493230}. Vector-based searches require numerous distance calculations and lookups across thousands of vectors, particularly with IVF-based methods \cite{jegou2010product,ge2013optimized,kalantidis2014locally}. Moreover, even compressed LMs impose significant overhead on mobile hardware \cite{sanh2019distilbert,jiao2019tinybert,sun2020mobilebert,xin2020deebert}. Since user satisfaction depends on a low Time to First Token (TTFT)—the interval from query submission to the first LLM token—this computational burden creates a self-reinforcing cycle that further degrades performance. For instance, Naive-RAG feeds an unoptimized 2K-token document (see Figure~\ref{fig:RAG}) directly to the LLM, and even though EdgeRAG and AdvancedRAG optimize memory usage and apply reranking, they still use the full 2K tokens. Consequently, retaining the complete documents as LLM input results in significant inference latency on mobile devices.

To handle these issues, we propose \textit{MobileRAG} that tackles these constraints through two key components. First, we present \textit{EcoVector}, a vector search method that partitions and partially loads large-scale data into smaller graph structures, reducing RAM usage, power consumption, and search latency. Second, we propose the \textit{SCR} method, which re-chunks retrieved documents, recalculates their similarity scores, and reconstructs the prompt by selecting only the most similar chunks, thereby significantly lowering inference time and energy use. Both components run on-device—eliminating network dependency and preserving privacy by avoiding external uploads—thereby enabling MobileRAG to achieve efficient retrieval and robust text generation under real mobile hardware constraints.

The contributions of this paper are as follows: 
\vspace*{-0.1cm}
\begin{itemize}
[labelindent=0em,labelsep=0.3em,leftmargin=*,itemsep=0em]
    \item We propose a mobile-tailored EcoVector indexing method using graph partitioning and partial loading for efficient memory and power management.
    \item We derive analytical models to estimate memory, latency, and power for vector indexing on mobile devices.
    \item We propose the Selective Content Reduction method with similarity-based reordering to lower sLM inference latency and power.
    \item We validate that MobileRAG fully supports offline operation, safeguarding privacy on mobile devices.
    \item We develop a MobileRAG Chat prototype for interactive search, summarization, and analysis.
    \vspace*{-0.1cm}
\end{itemize}

The rest of this paper is organized as follows: 
Section~\ref{sec:MobileRAGChat} presents the MobileRAG Method. Section~\ref{sec:EcoVector} describes EcoVector, and Section~\ref{sec:SCR} explains SCR. Section~\ref{sec:Experiments} shows experimental results. Section~\ref{sec:RelatedWork} reviews existing vector search and RAG Methods. Finally, Section~\ref{sec:Conclusion} concludes the paper with discussions.

\begin{figure*}[t]
    \centering
        \includegraphics[width=0.9\linewidth]{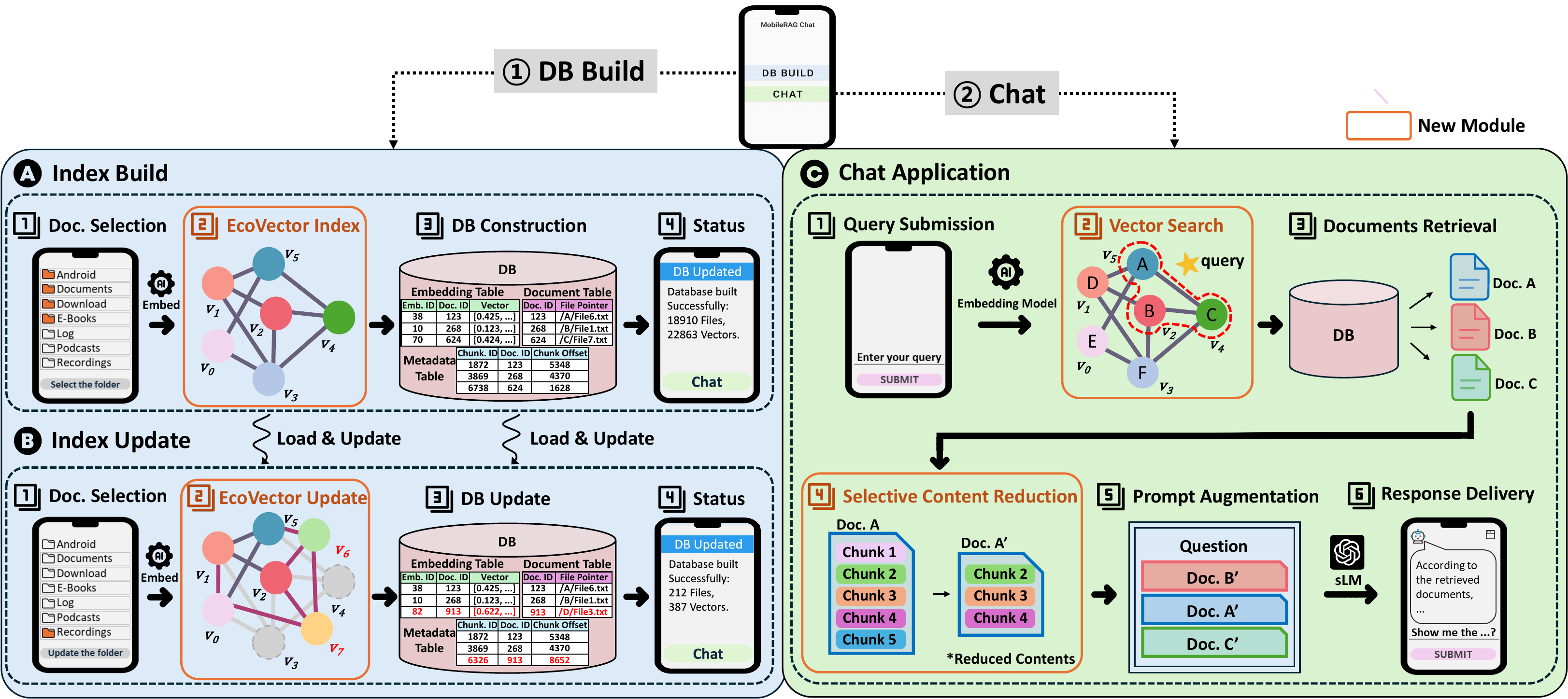}
            \vspace*{0.1cm}
        \label{fig:System_Overview}
    \caption{%
Overview of MobileRAG: on-device pipeline consisting of \textit{Index Build}, \textit{Index Update}, and \textit{Chat Application}.
    }
    \label{fig:System_Overview}
    \vspace*{-0.3cm}
\end{figure*}

\section{MobileRAG}\label{sec:MobileRAGChat}
MobileRAG operates fully on-device, locally constructing and maintaining its index with the \textit{EcoVector} index and the \textit{SCR} method for efficient retrieval and generation (Figure~\ref{fig:System_Overview}).

\subsection{Index Build}

Before any RAG operations can occur, the system must assemble the relevant documents and convert them into an indexable form. This \textit{Index Build} phase is divided into four main steps:

\vspace*{0.1cm}
\noindent\textbf{Document Selection:}
The user determines which documents stored on the mobile device should be included in \textit{MobileRAG}. These selected documents become the data source for subsequent retrieval and generation. The chosen documents are split into manageable chunks. Each chunk is then converted into a vector representation using the embedding model.

\vspace*{0.1cm}
    \noindent\textbf{EcoVector Index:}
Once the documents are embedded, the EcoVector algorithm indexes them to enable faster and more power-efficient retrieval (as described in Section~\ref{subsec:buildmethod}). The resulting EcoVector index is then used for efficient document retrieval during EcoVector updates and for Chat Application's vector search.

\vspace*{0.1cm}
\noindent\textbf{DB Construction:}
As illustrated in Figure~\ref{fig:System_Overview}, the system populates a local SQLite database with three main tables: the \textit{Embedding Table}, which stores vector embeddings with unique embedding IDs and corresponding document IDs (e.g., \textit{embedding ID $38$} links to \textit{document $123$} with \textit{vector $[0.425, \dots]$}); the \textit{Document Table}, which maps document IDs to file paths (e.g., \textit{document $123$} is stored at \textit{/A/File6.txt}), linking entries to disk files; and the \textit{Metadata Table}, which contains auxiliary data for chunk-level retrieval, such as chunk IDs and embedding offsets (e.g., \textit{chunk ID $1872$ }for \textit{document $123$} with \textit{offset $5348$}).

\vspace*{0.1cm}
\noindent\textbf{Status:}
The screen confirms successful database construction, displaying the number of indexed files and vectors (e.g., \textit{18,910 Files}, \textit{22,863 Vectors}). Users can directly tap the \textit{Chat} button at the bottom to proceed.

\subsection{Index Update}

Over time, the user may wish to add or remove documents from the existing index. MobileRAG's \textit{Index Update} process accommodates these needs without rebuilding the entire pipeline:

\vspace*{0.1cm}
\noindent\textbf{Document Selection:}
The user or application designates which new or obsolete documents should be added or removed. Newly added documents undergo the same chunking and embedding procedure. Deleted documents are flagged so that their embeddings can be purged from the database.

\vspace*{0.1cm}
\noindent\textbf{EcoVector Update:}
The system reuses the EcoVector index generated in the indexing phase and incrementally updates it upon insertions or deletions, as illustrated in the \textit{EcoVector Update }scenario in Figure~\ref{fig:System_Overview} (e.g., removing IDs \textit{$v_3$}, \textit{$v_4$}, and inserting IDs \textit{$v_5$}, \textit{$v_6$}), avoiding a full rebuild (detailed in Section~\ref{subsec:Update}).

\vspace*{0.1cm}
\noindent\textbf{DB Update:}
Once EcoVector completes its graph updates, the subsequent step is the \textit{DB update} process, as depicted in Figure~\ref{fig:System_Overview}. For example, when new vectors (e.g., embedding ID \textit{82}) and new documents (e.g., document ID \textit{913}) are inserted as shown in Figure~\ref{fig:System_Overview}, the DB update process adds these entries to the respective tables and removes outdated information to reflect recent changes accurately. The newly added embeddings are stored in the Embedding Table, document content is inserted into the Document Table, and the Metadata Table is updated to reflect accurate offsets and references.

\vspace*{0.1cm}
\noindent\textbf{Status:}
The screen confirms the database update, displaying the number of newly added files and vectors (e.g., \textit{212 Files}, \textit{387 Vectors}). Users can directly tap the \textit{Chat} button at the bottom to proceed.

\subsection{Chat Application}

Once the on-device database is established or updated, MobileRAG can be used to perform queries via a \textit{Chat} interface:

\vspace*{0.1cm}
\noindent\textbf{Query Submission:}
The user enters a natural language query through the MobileRAG UI. After pressing \textit{submit}, the query is embedded using the same model employed for document embeddings, ensuring consistency.

\vspace*{0.1cm}
\noindent\textbf{Vector Search:}
Using the embedded query, MobileRAG searches the local EcoVector index to retrieve the top-$k$ most relevant documents (or chunks). Section~\ref{subsec:Search} provides further details on this search mechanism.

\noindent\textbf{The Selective Content Reduction Method:}
Even after retrieving $k$ relevant documents, passing them all to an on-device sLM may be infeasible due to latency and power constraints. The SCR method addresses this by filtering documents, significantly reducing input size for the sLM without compromising retrieval accuracy (detailed in Section~\ref{sec:SCR}).

\vspace*{0.1cm}
\noindent\textbf{Prompt Augmentation and sLM Inference:}
By shrinking the total token count, we combine the condensed text and the original query into a final prompt, which is then presented to the on-device sLM.

\vspace*{0.1cm}
\noindent\textbf{Response Delivery:}
Finally, the sLM's output is displayed in the chat UI, providing the user with a context-rich, natural language response.

\begin{figure}[h]
\vspace*{-0.3cm}
  \includegraphics[width=\linewidth]{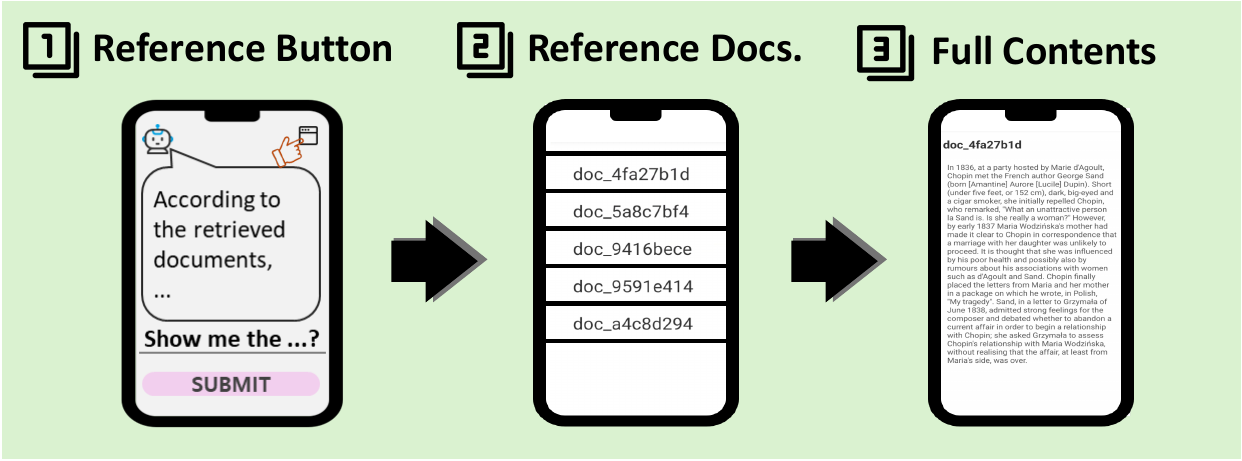}
  \hfill
\vspace*{-0.3cm}  
  \caption{Document References in Chat Application.}
  \label{fig:Reference}
\end{figure}

\vspace*{0.1cm}
On the Response Delivery page, the user can tap on a \textit{References} button to see which documents were used in generating the answer, as illustrated in Figure~\ref{fig:Reference}. Selecting any of these document titles reveals the full content of that document, enabling further review or validation of the sLM's sources.

\vspace{-0.3cm}
\section{EcoVector}\label{sec:EcoVector}
In this section, we present \textit{EcoVector}, covering its construction and search procedure, along with a theoretical analysis of its advantages in memory usage, search latency, and power consumption. We then introduce insertion and deletion methods that enable dynamic index updates while preserving EcoVector’s efficiency benefits.

\begin{figure}[h]
\vspace*{-0.2cm}
  \includegraphics[width=\linewidth]{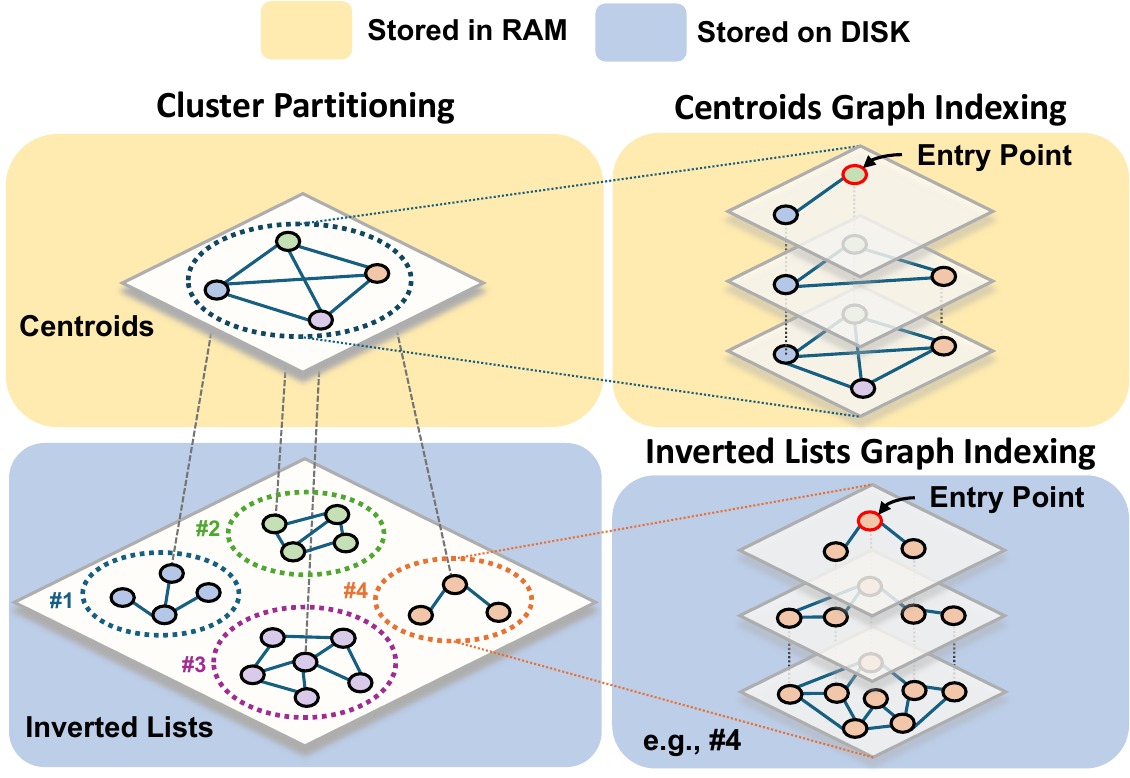}
  \caption{EcoVector Architecture.}
  \Description{EcoVector Indexing}
      \label{fig:ecovector}
            \vspace*{-0.2cm}
\end{figure}

\subsection{Build Method} \label{subsec:buildmethod}
The EcoVector index is constructed in four main stages as shown in Figure~\ref{fig:ecovector}: 

\subsubsection{Cluster Partitioning. }
The entire set of vectors are partitioned into clusters using an unsupervised clustering algorithm such as k-means. Figure~\ref{fig:ecovector} illustrates this clearly: here, vectors are divided into four distinct clusters (\#1--\#4) based on their embedding similarity. Each cluster is represented by a centroid (highlighted nodes), which serves as the representative embedding for that cluster. This partitioning process establishes the foundation for effectively reducing the search space during query processing. Given the limited RAM available on mobile devices, it is essential to support partial loading on a per-cluster basis rather than storing all vectors in memory. Moreover, the size of each inverted list remains manageable, allowing for more efficient control over CPU computations and memory usage during search.

\subsubsection{Construction of the Centroids Graph. }
Once clusters have been defined, an HNSW graph is constructed over the dataset consisting solely of the cluster centroids. Unlike traditional HNSW approaches that build a hierarchical graph over all vectors, EcoVector constructs the graph using only the representative vectors (centroids) of each cluster, resulting in a significantly reduced number of nodes. For instance, if there are on the order of tens of thousands of centroids, the HNSW structure can be maintained in RAM using only tens to a few hundred megabytes. On mobile devices, where CPU usage directly translates into power consumption, the centroids graph enables the approximation of high-dimensional distance calculations via efficient nearest neighbor search. 

\subsubsection{Construction of the Inverted Lists Graph. }
Next, an HNSW graph is built for the vectors contained within each cluster, forming what we refer to as the \textit{Inverted Lists Graph}, illustrated at the bottom of Figure~\ref{fig:ecovector}. Each cluster (\#1–\#4) independently forms its own graph with its embeddings. In a server environment, this massive graph might be entirely loaded into RAM to facilitate rapid search; however, on mobile devices, this is not feasible due to memory constraints. Therefore, EcoVector constructs an independent graph for each cluster and stores these inverted lists on disk (i.e., flash storage). Although building the inverted lists graph introduces additional indexing overhead, it significantly reduces the number of distance calculations required during query processing, thereby lowering CPU usage. 

\subsubsection{RAM-Disk Partitioned Storage. }
In the final stage, the centroids graph is kept in RAM, while the inverted lists graphs for each cluster are stored on disk. Since the number of centroids is relatively small compared to the entire dataset, their memory footprint is minimal and they are essential for every query—making in-RAM storage highly advantageous. Conversely, loading all inverted lists graphs into RAM simultaneously is impractical given mobile devices' limited memory; hence, these graphs are stored on disk and loaded dynamically only when required by a query. This strategy allows the device to selectively load data on a per-cluster basis and unload graphs after query processing, thereby freeing up RAM for other operations. On mobile devices, CPU computations tend to consume more power and generate more heat compared to disk I/O (detailed in Section~\ref{subsec:Power}). Thus, even if disk I/O increases, reducing CPU processing time is beneficial for overall power efficiency.

\subsection{Search Method} \label{subsec:Search}
After the EcoVector index is built, query search proceeds through three main stages: 

\subsubsection{Centroids Graph Search: }
When a query is received, the system first searches the Centroids Graph residing in RAM to identify the centroids that are closest to the query vector. Thanks to the hierarchical structure of HNSW, the system performs a $k$-ANNS instead of directly computing high-dimensional distances, thereby significantly reducing CPU processing time. As a result of this stage, a set of nearby centroids is determined, and the clusters they represent are selected for further processing.

\subsubsection{Loading Selected Cluster Graphs from Disk: }
From the clusters identified in the Centroids stage, only those deemed promising are loaded from disk into RAM. Given that mobile devices lack the capacity to load all clusters simultaneously, only the necessary clusters are selectively retrieved.

\subsubsection{Inverted Lists Graph Search: }
For each cluster graph loaded into RAM, the system searches for the actual data points that are closest to the query vector. The graph structure obviates the need for exhaustive distance computations between high-dimensional vectors, thereby shortening CPU processing times and reducing power consumption. Once the search within a cluster is complete, the corresponding graph is immediately unloaded to free up RAM. If necessary, the system then sequentially loads the next cluster and repeats the process. 

\subsection{Update Method} \label{subsec:Update}
After indexing, new data should be added or obsolete entries removed without rebuilding the entire graph.

\subsubsection{Insertion: } 
Algorithm~\ref{alg:insertPoint} describes the insertion procedure into the graph-based index, clearly illustrated by the \textit{EcoVector Update} scenario in Figure~\ref{fig:System_Overview}. Assume the existing graph comprises nodes \textit{$v_0$, $v_1$, $v_2$, $v_3$, $v_4$, and $v_5$} without any deletions.

When inserting new nodes (\textit{$v_6$} and \textit{$v_7$}), Algorithm~\ref{alg:insertPoint} first determines the suitable insertion level. It begins the search from the top-level entry point, progressively moving downward through the levels to identify an optimal position for the new nodes.

The algorithm then proceeds to expand candidates using the \textit{expandCandidates} function. Specifically, nodes closest to the insertion point are identified as initial candidates. For example, when inserting node \textit{$v_6$}, candidate nodes initially identified based on proximity criteria might include \textit{$v_2$, $v_3$, $v_4$,} and \textit{$v_5$}. Among these, the RobustPrune step selects only the most optimal neighbors by evaluating connectivity and proximity, resulting in node \textit{$v_6$} forming direct connections with nodes \textit{$v_2$} and \textit{$v_3$}, while candidates \textit{$v_4$} and \textit{$v_5$} are excluded.

Next, when inserting node \textit{$v_7$}, the algorithm identifies another set of candidate nodes, potentially including nodes \textit{$v_0$, $v_1$, $v_2$,} and the newly inserted node \textit{$v_6$}. After the RobustPrune step, which evaluates candidates based on connectivity strength and proximity, node \textit{$v_7$} finally establishes direct connections with nodes \textit{$v_0$, $v_2$,} and \textit{$v_6$}, achieving efficient integration into the existing graph structure.

Finally, Algorithm~\ref{alg:insertPoint} employs the \textit{connectTwoWay} function to ensure bidirectional connectivity, explicitly updating connections between the newly inserted nodes (\textit{$v_6$} and \textit{$v_7$}) and their selected neighbor nodes. This ensures robust integration of the inserted nodes into the existing hierarchical graph structure, effectively preserving overall search efficiency and structural integrity.
\vspace*{-0.2cm}

\begin{algorithm}[t]
\SetAlgoLined
\SetArgSty{textnormal}

\def\HiLi{\leavevmode\rlap{\hbox to \dimexpr\hsize+12pt\relax{\color{lightgray!55}\leaders\hrule height .8 \baselineskip depth .5ex\hfill}}}

\KwIn{\textit{id} /* node ID */, \textit{graph} /* HNSW graph */}
\KwOut{Updated \textit{graph}}

\SetKwProg{Fn}{Function}{:}{}
\Fn{insertPoint(\textit{id}, \textit{graph})}{
  \textit{vec} $\leftarrow$ reconstruct(\textit{id}); \\
  \HiLi \textit{lvl} $\leftarrow$ \textit{graph.levels}[\textit{id}]\;
  \If{$\textit{lvl} \le 0$}{
    \textit{lvl} $\leftarrow$ getRandomLevel($1.0/\log(\textit{maxM})$); \\
    \textit{graph.levels}[\textit{id}] $\leftarrow$ \textit{lvl}\;
  }
  \HiLi \textit{cur} $\leftarrow$ \textit{graph.entry\_point}\;
  \For{$\textit{l} \leftarrow \textit{graph.max}$ \textbf{downto} $\textit{lvl}+1$}{
    \Repeat{no improvement}{
      \ForEach{\textit{nb} in neighbors(\textit{cur}, \textit{l})}{
        \If{$\textit{nb} < 0$ or $\textit{is\_deleted}[\textit{nb}]$}{continue\;}
        \If{$dist(\textit{nb}, \textit{id}) < dist(\textit{cur}, \textit{id})$}{
          \textit{cur} $\leftarrow$ \textit{nb}\;
        }
      }
    }
  }
  \For{$\textit{l} \leftarrow \min(\textit{lvl}, \textit{graph.max})$ \textbf{downto} $0$}{
    \HiLi \textit{cand} $\leftarrow$ expandCandidates(\textit{cur}, \textit{id}, \textit{l}, \textit{ef})\;
    \HiLi \textit{fnbr} $\leftarrow$ robustPrune(\textit{cand}, \textit{alpha}, \textit{maxM}, \textit{id})\;
    \HiLi connectTwoWay(\textit{id}, \textit{fnbr}, \textit{l}, \textit{graph})\;
  }
  \textit{is\_deleted}[\textit{id}] $\leftarrow$ false\;
  \If{$\textit{lvl} > \textit{graph.max}$}{
    \textit{graph.max} $\leftarrow$ \textit{lvl}\;
    \textit{graph.entry\_point} $\leftarrow$ \textit{id}\;
  }
}
\caption{\textsc{Graph Insertion}}
\label{alg:insertPoint}
\end{algorithm}

\subsubsection{Deletion: }
Algorithm~\ref{alg:dd} describes the hierarchical deletion method used in graph-based indices, ensuring efficient maintenance of connectivity upon node removal. 

This deletion process is clearly illustrated by the \textit{EcoVector Update} scenario in Figure~\ref{fig:System_Overview}, where nodes \textit{$v_3$} and \textit{$v_4$} are removed. After selecting these nodes for deletion, Algorithm~\ref{alg:dd} first handles high-level graph adjustments, such as updating entry points and maximum levels, if necessary.

Next, the \textit{recNeighbors} procedure ensures that the remaining nodes remain well-connected post-deletion. Specifically, all links between deleted nodes (\textit{$v_3$, $v_4$}) and their neighbors (\textit{$v_0$, $v_1$, $v_2$, $v_5$}) are removed (represented as greyed-out links in Figure~\ref{fig:System_Overview}). Subsequently, a candidate set is constructed for each affected node to identify potential new neighbor connections, incorporating both existing neighbors and k-nearest nodes. The robust pruning heuristic then selects the best neighbors based on a combination of node distances and overall graph connectivity quality, restricting each node to a maximum of \textit{$M$} links to control graph density. For instance, in Figure~\ref{fig:System_Overview}, nodes \textit{$v_0$} and \textit{$v_1$}, previously indirectly connected via deleted nodes \textit{$v_3$} and \textit{$v_4$}, now establish a new connection to restore the connectivity.

Thus, the hierarchical deletion and neighbor reconnection approach systematically ensures both structural integrity and efficient index maintenance.

\begin{algorithm}[h]
\SetAlgoLined
\SetArgSty{textnormal}

\def\HiLi{\leavevmode\rlap{\hbox to \dimexpr\hsize+12pt\relax{\color{lightgray!55}\leaders\hrule height .8 \baselineskip depth .5ex\hfill}}}
\KwIn{\textit{id} /* node ID */, \textit{graph} /* HNSW graph */}
\KwOut{Updated \textit{graph}}
\SetKwProg{Fn}{Function}{:}{}
\Fn{Hierarchical\_Graph\_Deletion(\textit{label, graph})}{
    \HiLi \If{\textit{label} = \textit{graph.entry\_point}}{
        \textit{new\_entry\_point}, \textit{new\_max} $\leftarrow -1$\;
        \HiLi \If{\textit{new\_entry\_point} = -1}{
            \textit{graph.entry\_point} $\leftarrow -1$\;
            \textit{graph.max} $\leftarrow 0$\;
        }
        \Else{
            \textit{graph.entry\_point} $\leftarrow$ \textit{new\_entry\_point}\;
            \textit{graph.max} $\leftarrow$ \textit{new\_max}\;
        }
    }
    \ElseIf{\textit{graph.levels[label]} = \textit{graph.max}}{
       checkAndDecreaseMaxLevel()\;
    }
    \For{\textit{lvl} $\leftarrow 0$ \textbf{to} \textit{graph.levels[label]}}{
        \HiLi \textit{oldNeighbors} $\leftarrow$ neighbors of \textit{label} at level \textit{l}\;
        \HiLi recNeighbors(\textit{label}, \textit{graph}, \textit{oldNeighbors}, \textit{l})\;
    }
    \HiLi removePhysicalNode(\textit{label})\;
    \textit{is\_deleted[label]} $\leftarrow$ true\;
}
\caption{\textsc{Hierarchical Graph Deletion}}
\label{alg:dd}
\end{algorithm}


\subsection{Time and Space Complexities}\label{sec:time_space}
We theoretically compare EcoVector with IVF, IVFPQ, and their disk-based variants. Disk-based methods store inverted lists on disk to reduce memory usage.

    \vspace*{-0.1cm}
\subsubsection{Analysis for Memory Usage:} \label{subsec:Memory}
We categorize memory usage into three components.

\begin{table}[b]
  \vspace*{-0.1cm}
  \setlength{\tabcolsep}{1.0pt}
  \begin{tabular}{@{}ll@{}}
    \toprule
    \textbf{Algorithm} & \textbf{Memory Usage Expression} \\ 
    \midrule
    IVF & \( N_c \cdot 4d + 8N + N\cdot4d \) \\[1ex]
    IVFPQ & \( N_c\cdot4d + 8N + N\Bigl(M_{\text{pq}}\cdot\frac{nbits}{8}\Bigr) + 2^{nbits}\cdot4d \) \\[1ex]
    HNSW & \( N\cdot4d + 4N\cdot \frac{M}{1-p_0} \) \\[1ex]
    HNSWPQ & \( N\Bigl(M_{\text{pq}}\cdot\frac{nbits}{8}\Bigr) + 4N\cdot  4\frac{M}{1-p_0} + 2^{nbits}\cdot4d \) \\[1ex]
    IVF-DISK & \( N_c\cdot4d + 8N + 4d \) \\[1ex]
    IVFPQ-DISK & \( N_c\cdot4d + 8N + M'_{\text{pq}}\cdot\frac{nbits}{8} + 2^{nbits}\cdot4d \) \\[1ex]
    IVF-HNSW & \( 4N_c\Bigl(d+\frac{M'}{1-p_0}\Bigr) + 8N + 4d \) \\[1ex]
    EcoVector & \( 4N_c\Bigl(d+\frac{M'}{1-p_0}\Bigr) + 8N + 4\Bigl(d+\frac{M'}{1-p_0}\Bigr) \) \\
    \bottomrule
  \end{tabular}
      \vspace*{0.1cm}
    \caption{Memory Usage Expressions.}
      \label{tab:memory-expressions}
      \vspace*{-0.2cm}
\end{table}

\noindent\textbf{Clustering.} Methods that partition the dataset store \(N_c\) centroids of dimension \(d\), inverted lists assigning each of the 
\(N\) vectors to a cluster, and 8-byte IDs:
\vspace*{-0.2cm}
\[
  \text{Mem}_{\text{Clustering}} \approx N_c \cdot d \cdot 4 \;+\; N \cdot 8 + N \cdot d \cdot 4
\vspace*{-0.2cm}
\]
\noindent\textbf{Graph.} Graph-based approaches (e.g., HNSW) maintain full embeddings and neighbor links. All $N$ vectors are stored in $d$ dimensions (4 bytes each). If each vector has up to $M$ neighbors and the probability of having a level $\geq l$ is $p_0^l$ with $p_0 = \frac{1}{\ln(M)}$, the memory required at level \(l\) is
    \vspace*{-0.1cm}
\[
  \text{Mem}_{\text{neighbors}}(l) = N  \cdot M \cdot 4 \cdot p_0^l.
\]

Summing over all levels (from \(l=0\) to \(L_{\max}-1\)) and approximating with an infinite geometric series (assuming \(L_{\max}\) is large enough), we obtain
    \vspace*{-0.1cm}
\[
  \text{Mem}_{\text{neighbors}} \approx N \cdot M \cdot 4 \cdot \frac{1}{1-p_0}
\]
    \vspace*{-0.1cm}
Thus, the total graph memory is given by
    \vspace*{-0.1cm}
\[
  \text{Mem}_{\text{Graph}} \approx N \cdot d \cdot 4 \;+\; N \cdot M \cdot 4 \cdot \frac{1}{1-p_0}
\]

\vspace*{-0.1cm}
\noindent\textbf{PQ.} For compression, each vector is split into $M_{\text{pq}}$ sub-vectors, each encoded in $nbits$ bits. Hence, storing $N$ compressed vectors the per-vector memory requirement is
\[
  \text{Mem}_{\text{PQ, per vector}} = N \cdot \Bigl(M_{\text{pq}} \cdot \frac{nbits}{8}\Bigr)
\]

The codebook contains \(2^{nbits}\) codewords, thus, the overall PQ Memory is 
\vspace*{-0.1cm}
\[
  \text{Mem}_{\text{PQ}} \approx N \cdot \Bigl(M_{\text{pq}} \cdot \frac{nbits}{8}\Bigr) \;+\; 2^{nbits} \cdot d \cdot 4
\]

\vspace*{-0.1cm}
\noindent Table~\ref{tab:memory-expressions} summarizes these expressions for each baseline.

\subsubsection{Analysis for Search Latency:}\label{subsec:Search}

Search latency on mobile devices comprises CPU-based processing ($t_s$) and disk I/O delays ($t_d$).
\[
T_{\text{search}} \;=\; t_s \;+\; t_d
\]

\noindent\textbf{CPU-Based Search Time $(t_s)$.}
Let $n_{\text{search}}$ be the total number of distance computations, and $t_{op}$ the time per distance. Then
  \vspace*{-0.01cm}
\[
t_s = n_{\text{search}} \,\cdot\, t_{op}
\quad\text{where}\quad
t_{op} = \frac{\# \,\text{of CPU cycles}}{\text{CPU clock frequency}}
\]

In our setting, one distance computation requires about 500 CPU cycles for a
128-dimensional vector; at 2.4\,GHz, this yields \(t_{op} \approx 1.94 \cdot 10^{-4}\) ms \cite{sre_google}.  The term $n_{\text{search}}$ varies by algorithm. Clustering-based approaches first compare the query against all $N_c$ centroids, then probe $n_P$ clusters (about $N_c + n_P \cdot \tfrac{N}{N_c}$ operations), while graph-based methods often incur $ef \cdot M$ operations (where $ef$ is the expansion factor). Table~\ref{tab:time-complexity} summarizes the respective formulas for each baseline method.

\begin{table}[t]
  \setlength{\tabcolsep}{1.0pt}
  \begin{tabular}{@{}ll@{}}
    \toprule
    \textbf{Algorithm} & \textbf{Search Time Expression} \\ 
    \midrule
    IVF & \( N_c +  n_P \cdot \frac{N}{N_c} \) \\[1ex]
    IVFPQ & \( N_c + n_P \cdot \frac{N}{N_c} \cdot \frac{M_{\text{pq}}}{d} \cdot \frac{{n_{bits}}}{8}  + 2^{n_{bits}}\) \\[1ex]
    HNSW & \( ef_H \cdot M_h \) \\[1ex]
    HNSWPQ & \( ef_H \cdot M_h \cdot \frac{M_{\text{pq}}}{d} \cdot \frac{{n_{bits}}}{8} + 2^{n_{bits}} \) \\[1ex]
    IVF-DISK & \( N_c + n_P \cdot \frac{N}{N_c} \) \\[1ex]
    IVFPQ-DISK & \( N_c + n_P \cdot \frac{N}{N_c} \cdot \frac{M_{\text{pq}}}{d}\cdot \frac{{n_{bits}}}{8}  + 2^{n_{bits}}\) \\[1ex]
    IVF-HNSW & \( ef_c \cdot M' + n_P \cdot \frac{N}{N_c} \) \\[1ex]
    EcoVector  & \( ef_c \cdot M' + n_P \cdot ef_L \cdot M' \) \\
    \bottomrule
  \vspace*{-0.3cm}
  \end{tabular}
    \caption{Search Latency Expressions for Baselines.}
      \label{tab:time-complexity}
    \vspace*{-0.8cm}
\end{table}

\noindent\textbf{Disk I/O Time $(t_d)$.}
If $n_{\text{seek}}$ denotes the number of random file-pointer moves, $T_{\text{seek}}$ the seek time, $T_{\text{cmd}}$ a per-access command overhead, $n_{\text{byte}}$ the data in bytes per load, and $T_{\text{transfer}}$ the transfer time per byte, then
\vspace{-0.4cm}

\begingroup
  \setlength{\abovedisplayskip}{5pt}
  \setlength{\abovedisplayshortskip}{0pt}
  \setlength{\belowdisplayskip}{5pt}
  \setlength{\belowdisplayshortskip}{0pt}
\[
t_d \;=\; n_{\text{seek}}\;\cdot\;\Bigl(T_{\text{seek}} \;+\;T_{\text{cmd}} \;+\; n_{\text{byte}}\, \cdot\ T_{\text{transfer}}\Bigr)
\] \endgroup

In our setting, $n_{\text{seek}} = n_{P}$. The values for $T_{\text{seek}}$ and $T_{\text{transfer}}$ are derived from official UFS~4.0 specifications, while $T_{\text{cmd}}$ is empirically measured to capture real operating conditions \cite{sre_google, lucas2022ditis}. Because UFS~4.0 supports up to $40{,}000$~IOPS at $2800$\,MB/s, we set $T_{\text{seek}} \approx 0.025$\,ms, $T_{\text{cmd}}=0.015$\,ms, and $T_{\text{transfer}} \approx 3.6\times10^{-7}$\,ms/Byte \cite{samsung_ufs, carroll2010analysis}. We load one inverted list at a time and release it after use, so $n_{\text{byte}}$ corresponds to the memory footprint of a single list. Exact values depend on device specifics and OS-level optimizations.

\subsubsection{Analysis for Power Consumption: } \label{subsec:Power}
Smartphone batteries typically maintain a nearly constant voltage \(V\) (e.g., 3.8 –4.2\,V), so the total energy \(E\) consumed per search is:
\begingroup
  \setlength{\abovedisplayskip}{5pt}
  \setlength{\abovedisplayshortskip}{0pt}
  \setlength{\belowdisplayskip}{5pt}
  \setlength{\belowdisplayshortskip}{0pt}
\[
E = \int_{t_0}^{t_1} P(t)\, dt = \int_{t_0}^{t_1} \bigl[I(t) \cdot V\bigr]\, dt \approx V \cdot \Bigl[I(t_s)\cdot\,t_s + I(t_d)\cdot\,t_d\Bigr]
\] \endgroup
where \(t_s\) and \(t_d\) denote the CPU computation time and disk I/O time, respectively and are computed in Section~\ref{subsec:Search}. In our setting, we have determined that \(I(t_s) \approx 2300\,\mu\text{A}\) and \(I(t_d) \approx 800\,\mu\text{A}\). Thus, CPU-bound operations consume significantly more power than disk-bound operations \cite{anandtech, carroll2010analysis}.

From Sections~\ref{subsec:Memory}--\ref{subsec:Power}, IVF-DISK achieves the smallest RAM footprint by offloading most index content to disk, while EcoVector adds minimal overhead for intra-cluster HNSW and remains nearly as compact. By partitioning the dataset and building compact per-cluster graphs, EcoVector reduces the number of distance computations enough to counteract its disk-loading overhead, thereby achieving the fastest theoretical query times and lowest power usage among the baselines.

\section{Selective Content Reduction}\label{sec:SCR}
\begin{figure}[t]
    \centering
        \includegraphics[width=0.8\linewidth]{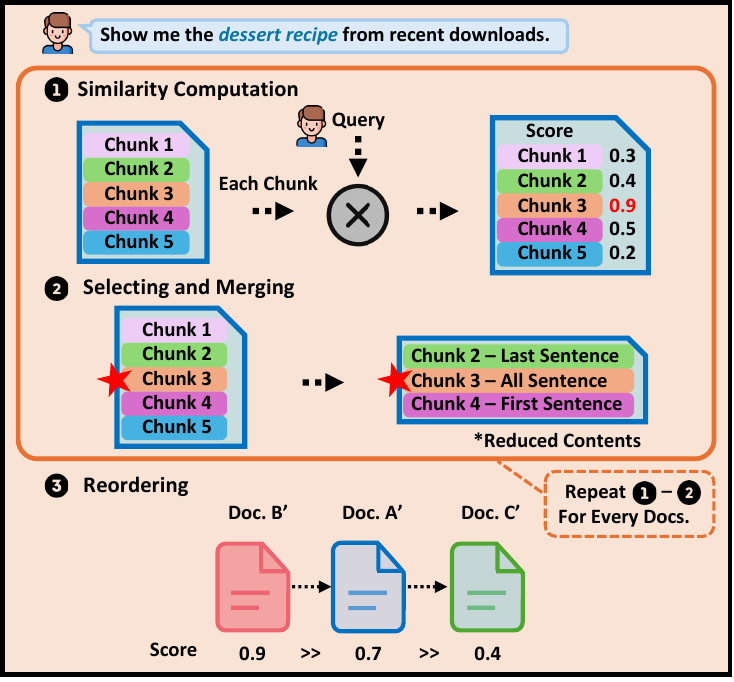}
        \caption{Overview of the SCR Method.}
    \label{fig:SCR_Overview}
    \vspace*{-0.2cm}
\end{figure}

A key challenge in mobile RAG is minimizing input size to the sLM. To address this, we introduce the SCR method, applied at the post-retrieval stage to reduce sLM input size, latency, and power consumption. Figure~\ref{fig:SCR_Overview} illustrates the overall architecture of MobileRAG incorporating the SCR method, which operates in three steps: \textit{Similarity Computation, Selecting and Merging, ReOrdering}.

Consider the scenario in Figure~\ref{fig:SCR_Overview} with query:
\vspace*{-0.2cm}

\noindent\textcolor{myred}{\rule{\columnwidth}{0.5pt}}
\textcolor{myred}{Query:} \textit{Show me the \textcolor{myblue}{dessert recipe} from recent downloads}.\\[-0.8ex]
\noindent\textcolor{myred}{\rule{\columnwidth}{0.5pt}}

\noindent Also, the initial retrieval stage has produced three documents—Documents A, B, and C. In what follows, we illustrate how the SCR method is applied to these retrieved documents through the three steps:

\noindent\textbf{{Step~1: Similarity Computation:}}
Instead of sending full initially retrieved documents directly into the sLM, each document is first split into individual sentences. Subsequently, sliding windows of a fixed size \textit{sliding\_window\_size} (e.g., three to five sentences per window) are generated with a predetermined step size \textit{overlap\_size}, creating overlapping segments. For example, if a document comprises five sentences, overlapping windows such as (sentences 1--3, 2--4, 3--5, etc.) are produced. The SCR method performs sentence-level segmentation only on the documents already filtered through initial similarity search, thereby minimizing overhead.

For example, consider a document (\textcolor{mygreen}{\textit{Doc. B}}) divided into five chunks as follows:
\vspace*{-0.2cm}

\noindent\textcolor{myred}{\rule{\columnwidth}{0.5pt}}
\textcolor{myred}{Chunk1:} "The Tiramisu dessert \textit{\textcolor{myblue}{originated}} in Italy ... " \\
\textcolor{myred}{Chunk2:} "An interesting \textit{\textcolor{myblue}{historical note}} about Tiramisu ... " \\
\textcolor{myred}{Chunk3:} "\textit{\textcolor{myblue}{Recipe}} of the Tiramisu includes cheese ... " \\
\textcolor{myred}{Chunk4:} "\textit{\textcolor{myblue}{The price}} of a single slice of Tiramisu can vary ... " \\
\textcolor{myred}{Chunk5:} "\textit{\textcolor{myblue}{Many cafés}} now offer Tiramisu for pick-up ... "\\[-0.8ex]
\noindent\textcolor{myred}{\rule{\columnwidth}{0.5pt}}

Here, Chunk1 discusses the origin, Chunk2 offers historical context, Chunk3 directly provides the recipe, Chunk4 details pricing, and Chunk5 mentions availability.

For each chunk, an embedding is recalculated, and its similarity to the query embedding is measured. Although performing sentence-level embedding and distance calculations entails additional CPU computations, these operations are considerably less demanding in terms of latency and power compared to running a full sLM over extensive text.

In our example, the scores are:
\vspace*{-0.2cm}

\noindent\textcolor{myred}{\rule{\columnwidth}{0.5pt}}
\textcolor{myred}{Query:} \textit{Show me the \textcolor{myblue}{dessert recipe} from recent downloads}.
\textcolor{myred}{Chunk1:} 0.3 (Discusses origin), \\
\textcolor{myred}{Chunk2:} 0.4 (Provides historical context),\\
\textcolor{myred}{Chunk3:} \textbf{0.9} (Contains the \textit{\textcolor{myblue}{Tiramisu recipe}}),\\
\textcolor{myred}{Chunk4:} 0.5 (Mentions pricing details),\\
\textcolor{myred}{Chunk5:} 0.2 (Focuses on store availability) \\[-0.8ex]
\noindent\textcolor{myred}{\rule{\columnwidth}{0.5pt}}

This process clearly indicates that Chunk3 is the most relevant to the recipe query. In contrast, Chunk1 and Chunk5 are found to be less relevant to the query.

\vspace*{0.1cm}
\noindent\textbf{{Step~2: Selecting and Merging:}}
From the sliding windows, the top-1 windows based on similarity scores are selected (per retrieved contents). To ensure that the contextual flow is preserved, \textit{context\_extension\_size} sentences are appended to both the beginning and the end of each selected window. Consequently, rather than feeding the entire content of all retrieved documents into the sLM, only the most relevant and contextually cohesive segments are merged. 

In our example, Chunk3 is chosen as the primary segment because it directly contains the recipe, while Chunks 2 and 4 are selected as the adjacent segments to Chunk3. To maintain contextual flow, we also incorporate the merged document (referred to as \textcolor{mygreen}{\textit{Doc B'}}) with clear source attribution (assume $\textit{context\_extension\_size} = 1$):
\vspace*{-0.2cm}

\noindent\textcolor{myred}{\rule{\columnwidth}{0.5pt}}
[\textcolor{myred}{From Chunk2} - last sentence]: "Then let's jump into how ..."\newline
[\textcolor{myred}{From Chunk3}]: "\textcolor{myblue}{\textit{Recipe of the Tiramisu}} includes cheese ..."\newline
[\textcolor{myred}{From Chunk4} - first sentence]: "The price of a single ..." \\[-0.8ex]
\noindent\textcolor{myred}{\rule{\columnwidth}{0.5pt}}

Then the final merged document appears as:
\vspace*{-0.2cm}

\noindent\textcolor{myred}{\rule{\columnwidth}{0.5pt}}
\textcolor{mygreen}{\textit{Doc. B':}} "Then let's jump into how to make Tiramisu. \textcolor{myblue}{\textit{Recipe of the Tiramisu}} includes cheese ... The price of a single slice of Tiramisu can vary depending on location" \\[-0.8ex]
\noindent\textcolor{myred}{\rule{\columnwidth}{0.5pt}}

Now, we obtain a more condensed version of the original \textcolor{mygreen}{\textit{Doc. B}}, referred to as \textcolor{mygreen}{\textit{Doc. B'}}. Although condensed, it still encompasses all the document content relevant to the query. Moreover, the process from \textit{Chunking} to \textit{Selecting and Merging} is applied to all retrieved documents.

\vspace*{0.1cm}
\noindent\textbf{Step~3: Reordering:}
After obtaining the condensed document, we reorder the individual documents according to their highest similarity scores. Specifically, each document is reordered based on its highest similarity score among its chunks. Suppose Document B' (which includes the recipe) has the highest relevance, followed by Document A' and then Document C'. We then reorder them as follows:
\vspace*{-0.2cm}

\noindent\textcolor{myred}{\rule{\columnwidth}{0.5pt}}
Original Sequence: \textcolor{mygreen}{\textit{Doc. A}} → \textcolor{mygreen}{\textit{Doc. B}} → \textcolor{mygreen}{\textit{Doc. C}} \\
Reordered Sequence: \textcolor{mygreen}{\textit{Doc. B'}} → \textcolor{mygreen}{\textit{Doc. A'}} → \textcolor{mygreen}{\textit{Doc. C'}} \\[-0.8ex]
\noindent\textcolor{myred}{\rule{\columnwidth}{0.5pt}}

This reordering acts as a Re-Ranker, enhancing the overall retrieval quality and thus improving the final performance of the RAG system.

\section{Experiments} \label{sec:Experiments}
\subsection{Experimental Setup}

On the hardware side, we used a Galaxy S24, which has hardware specifications representative of a mid-tier smartphone: 8\,GB of RAM, an Exynos 2400 CPU, a 4000\,mAh battery, and the Android~14 operating system~\cite{kawano2014ilsvrc,qu2025mobile}.

In our experiments, we employ both image and text ANNS datasets to evaluate EcoVector, while also testing the SCR method with three QA benchmarks. The image domain relies on SIFT~\cite{aumuller2020ann}, and text-based vectors come from NYTimes~\cite{aumuller2020ann}. For the MobileRAG setting, we use SQuAD~\cite{rajpurkar2016squad}, HotpotQA \\ ~\cite{yang2018hotpotqa}, and TriviaQA~\cite{joshi2017triviaqa}, which respectively encompass real-world user queries, multi-hop reasoning, and factoid-style QA. Table~\ref{tab:dataset_overview} summarizes all datasets and their dimensions.

\begin{table}[h]
  \vspace*{-0.2cm}
\caption{Summary of the datasets.}
  \vspace*{-0.2cm}
\begin{tabular}{l l r r r}
\toprule
\textbf{Type} & \textbf{Dataset} & \textbf{Base Vectors} & \textbf{Queries} & \textbf{Dim.} \\
\midrule
\multirow{2}{*}{ANNS} 
    & SIFT    & 1,000,000 & 10,000 & 128 \\
    & NYTimes & 290,000   & 10,000 & 256 \\
\midrule
\multirow{3}{*}{QA}
    & SQuAD    & 87,599  & 1,000 & 384 \\
    & HotpotQA & 90,447  & 7,405  & 384 \\
    & TriviaQA & 96,000  & 1,000  & 384 \\
\bottomrule
\end{tabular}
  \vspace*{-0.2cm}
\label{tab:dataset_overview}
\end{table}

\subsection{Evaluation of EcoVector} \label{subsec:EcoVector}

\subsubsection{Analysis for Memory Usage:} 
\begin{figure}[h]
  \vspace*{-0.2cm}
  \includegraphics[width=\linewidth]{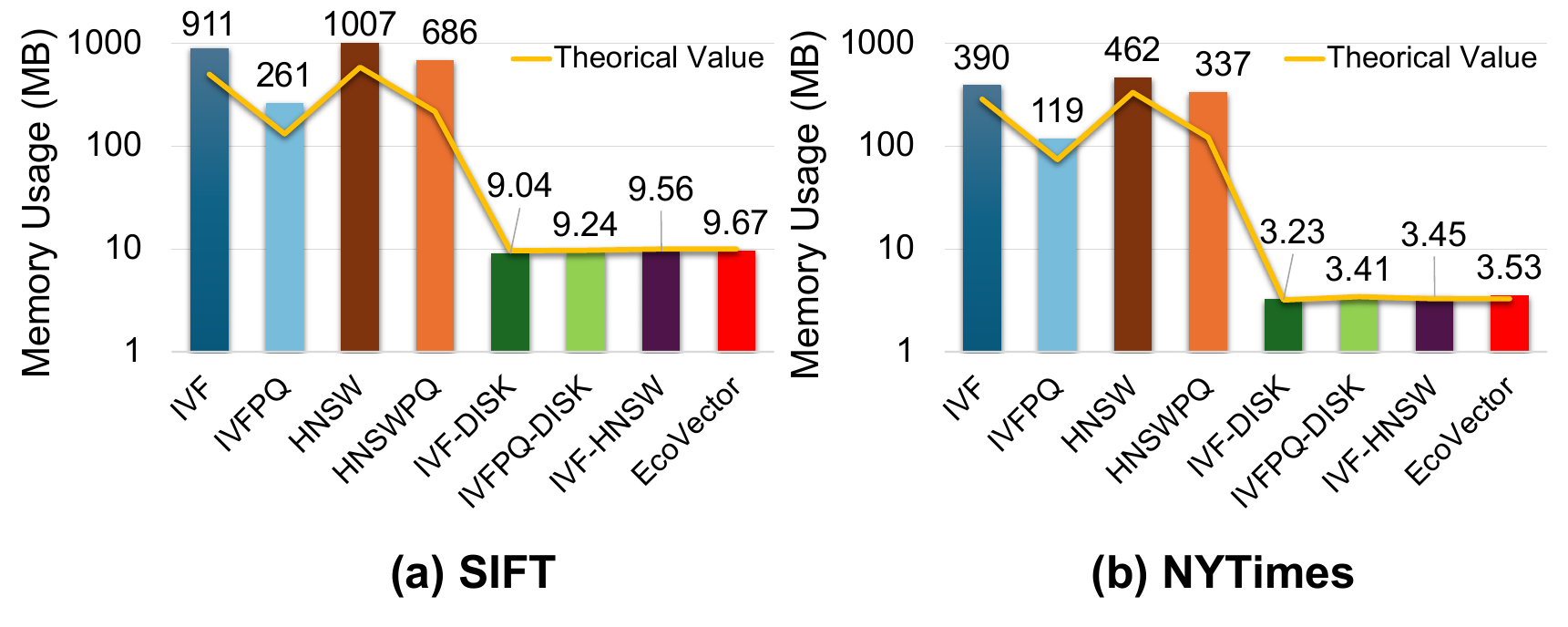}
  \caption{Memory Usage.}
    \label{fig:Memory}
      \vspace*{-0.5cm}
\end{figure}

Figure~\ref{fig:Memory} presents both the actual memory usage and the theoretical values from Section~\ref{sec:time_space} for the SIFT and NYTimes datasets. As shown, the disk-based methods (IVF-DISK, IVFPQ-DISK, IVF-HNSW, and EcoVector) all exhibit similar memory footprints, which means adding a graph index (e.g., in IVF-HNSW, EcoVector) does not significantly inflate memory consumption relative to IVF-DISK or IVFPQ-DISK; this is explained by the relatively small size of the per-cluster graphs, which on average comprise only 200–300 data points per cluster (Figure~\ref{fig:Distribution}a). In other words, although maintaining the graph structure requires additional memory, the minor overhead of storing multiple small cluster graphs proves worthwhile, especially given the marked reduction in search latency—a trade-off discussed further in our later analysis.

\subsubsection{Analysis for Search Latency:} 
\begin{figure}[t]
  \centering
  \includegraphics[width=\linewidth]{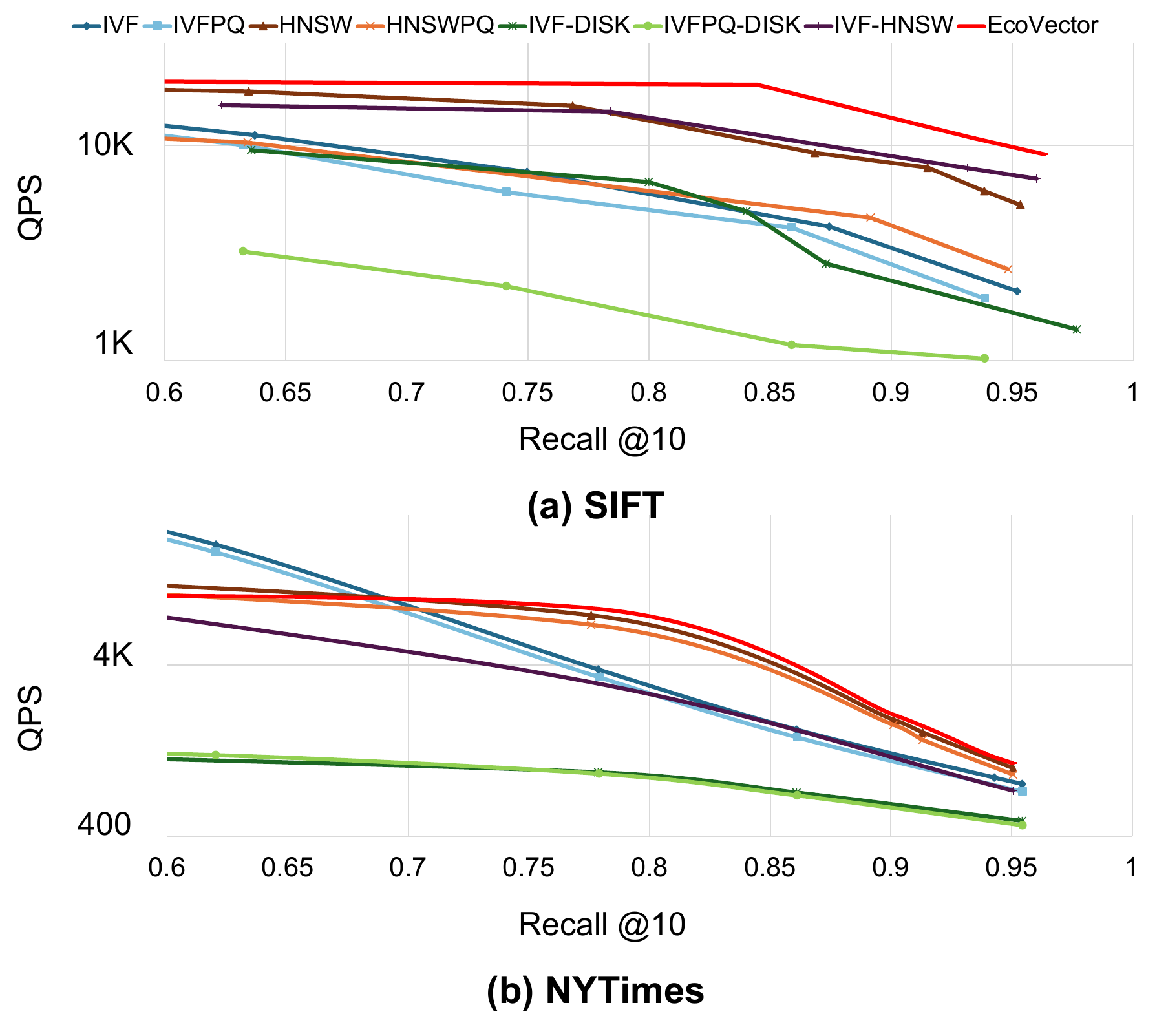}
    \vspace*{-0.4cm}
  \caption{Comparison of Recall and QPS.}
    \label{fig:QPS}
    \vspace*{-0.2cm}
    
\end{figure}

Figure~\ref{fig:QPS} reports Recall versus Queries Per Second (QPS) on the SIFT and NYTimes datasets. Despite incurring some disk I/O, EcoVector achieves the fastest query speeds overall by leveraging IVF-like clustering to filter candidates and then applying centroids and inverted-lists graph index. This combined effect more than compensates for disk overhead, granting EcoVector a clear advantage in end-to-end latency at a fixed recall level.

\begin{figure}[b]
  \centering
  \includegraphics[width=\linewidth]{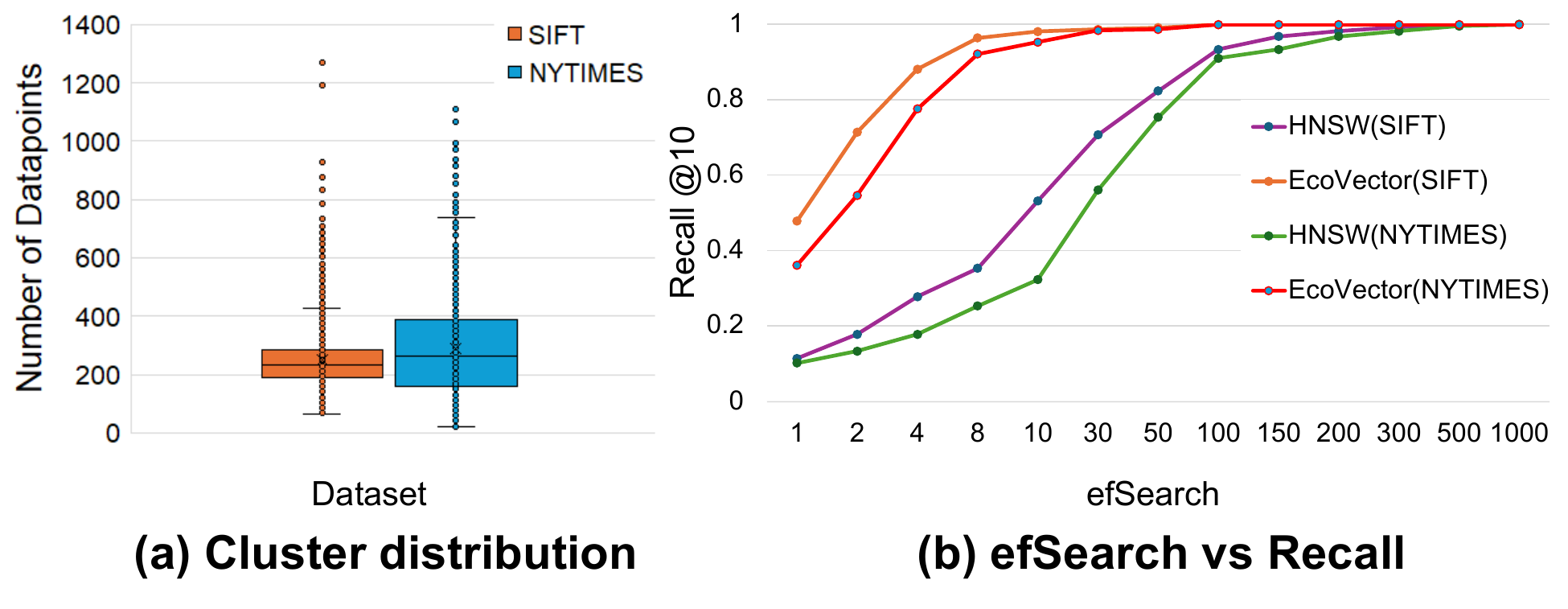}
  \caption{Cluster Distribution and Search Width.}
    \label{fig:Distribution}
          \vspace*{-0.4cm}
\end{figure}

Figure~\ref{fig:Distribution}a explains how EcoVector exploits multiple small graphs instead of a single massive structure (as in one-graph HNSW). Consequently, Figure~\ref{fig:Distribution}b indicates that these methods achieve high recall at a much smaller efSearch width than HNSW. This smaller search width effectively offsets disk I/O costs, leading to lower overall search latency—and, in turn, helping to minimize power consumption on mobile devices.

\subsubsection{Analysis for Power Consumption:}
\begin{figure}[h]
  \vspace*{-0.2cm}
  \includegraphics[width=\linewidth]{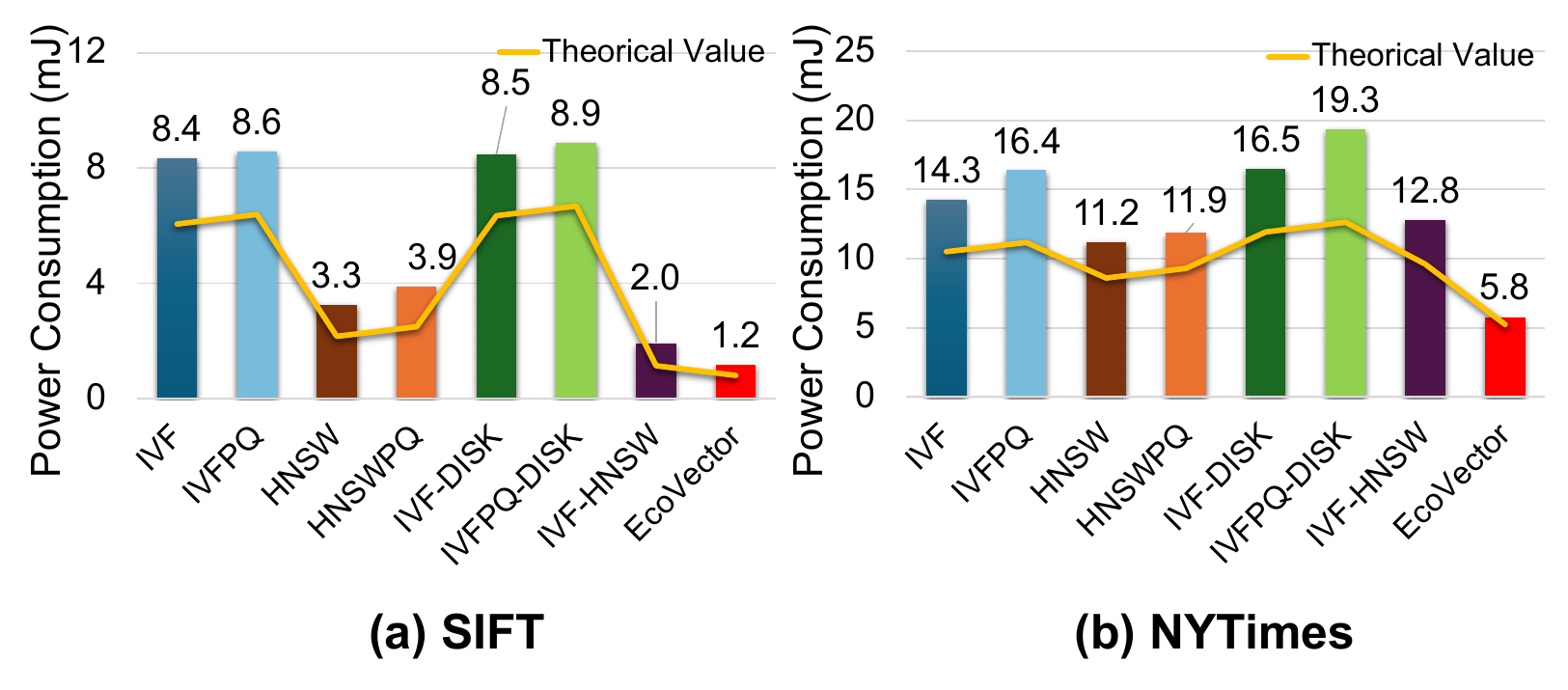}
  \caption{Power Consumption.}
    \label{fig:Power}
      \vspace*{-0.2cm}
\end{figure}

Figure~\ref{fig:Power} further confirms that EcoVector consumes noticeably less energy than its baselines. The primary reason is EcoVector’s strategy of subdividing the dataset into small cluster-based graphs (as shown in Figure~\ref{fig:Distribution}a), thus reducing CPU-based distance computations—the dominant factor in mobile power usage. While additional disk reads occur, the decrease in CPU-intensive tasks provides a net power advantage.

\subsubsection{Analysis for Update Latency:} 
\begin{figure}[h]
  \vspace*{-0.2cm}
  \includegraphics[width=\linewidth]{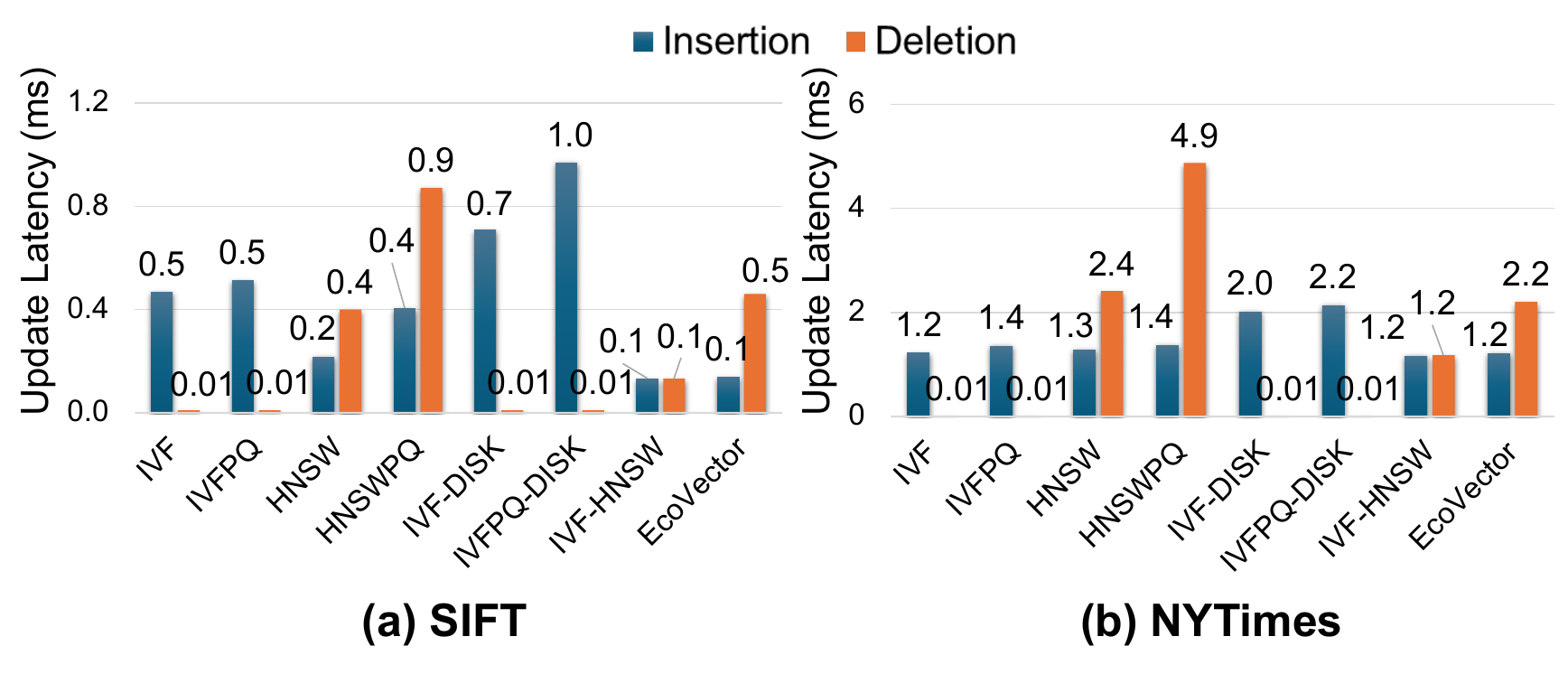}
    \vspace*{-0.4cm}
  \caption{Insertion and Deletion Latency.}
    \label{fig:UPDATE}
    \vspace*{-0.2cm}
\end{figure}

Figure~\ref{fig:UPDATE} presents insertion and deletion latencies for the SIFT and NYTimes datasets. EcoVector shows moderate insertion overhead by confining updates to small per-cluster graphs and achieves efficient deletion via minimal re-linking, thereby demonstrating balanced update performance.

Pure IVF systems attain very low update latency by keeping all data in RAM, which allows deletions to be executed by directly removing list entries. Similarly, IVF-HNSW benefits from simple inverted lists despite using HNSW for centroids. In contrast, EcoVector’s graph-based update mechanism requires identifying nodes and updating multiple links, leading to higher latency. This overhead is offset by gains in memory efficiency and power consumption: by localizing updates within compact per-cluster graphs, EcoVector minimizes unnecessary data movement and reduces memory footprint, while the limited re-linking confines computational effort, thereby enhancing power efficiency.

\subsubsection{Analysis for varying numbers of centroids:}  
\begin{figure}[t]
  \centering
  \includegraphics[width=\linewidth]{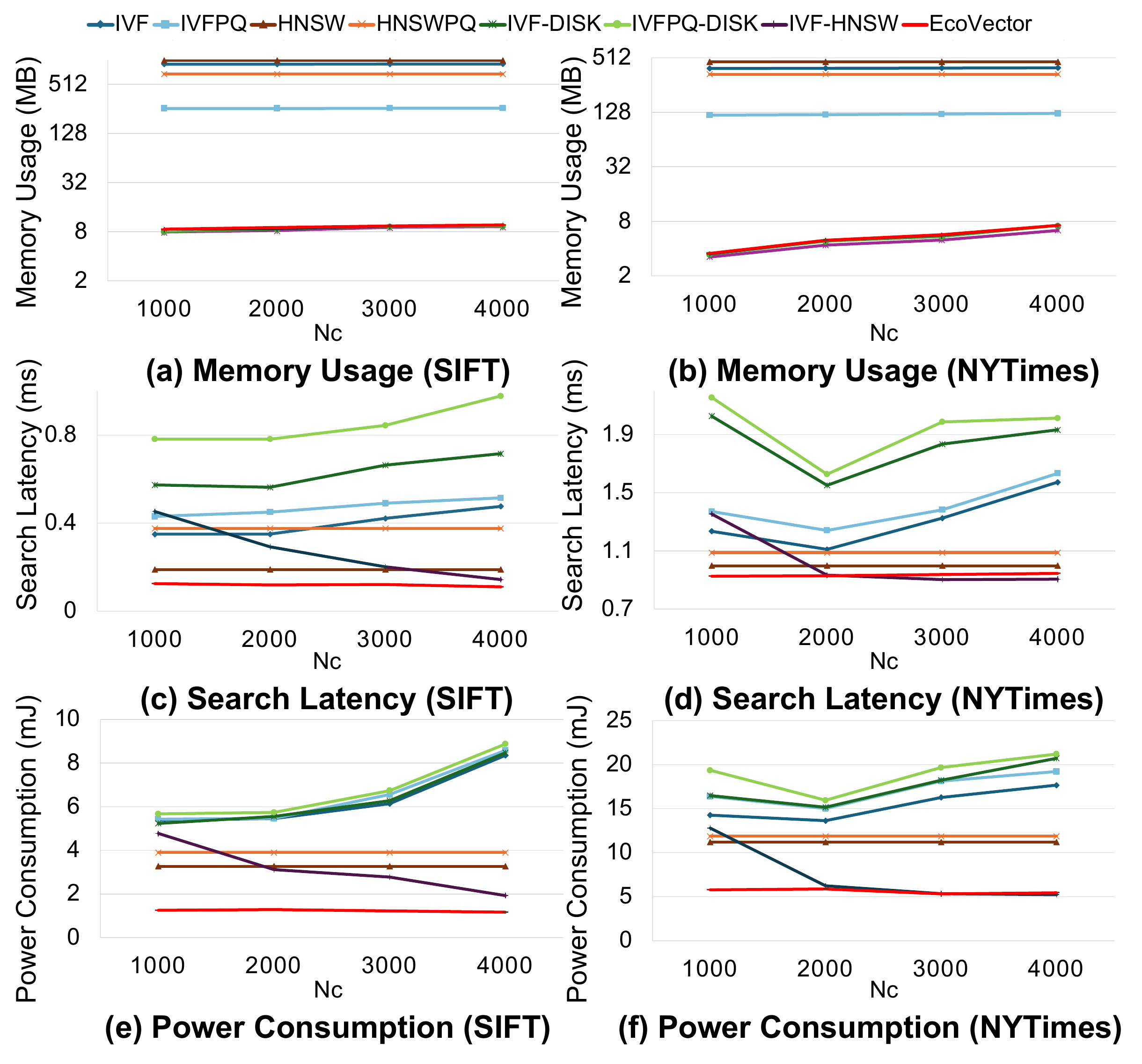}
    \vspace*{-0.3cm}
  \caption{Memory Usage, Search Latency, and Power Consumption on Various Cluster Sizes.}
    \label{fig:Experimental_Nc}
    \vspace*{-0.4cm}
\end{figure}

Figure~\ref{fig:Experimental_Nc} shows that on SIFT, increasing $N_c$ leads to a modest rise in EcoVector’s memory usage and a slight reduction in both search latency and power consumption. The changes remain small due to SIFT’s relatively low dimensionality. In contrast, on the NYTimes dataset—where dimensionality is higher—the memory usage grows more noticeably as $N_c$ increases, while search latency and power consumption follow a similar downward trend but also exhibit only marginal improvement.

\subsection{Evaluation of the SCR Method}
For both queries and documents, we employ the GTE-Small embedding model, which contains approximately 33 million parameters~\cite{li2023towards}. Additionally, we utilize the BERTSUM model as a compressor to further reduce the content size before inference~\cite{liu2019text}. For generation tasks, we utilized Qwen2.5 0.5B, 1.5B and Deepseek-r1 1.5B~\cite{yang2024qwen2,guo2025deepseek}. These models were executed via Ollama.

\begin{figure}[t]
  \centering
  \includegraphics[width=\linewidth]{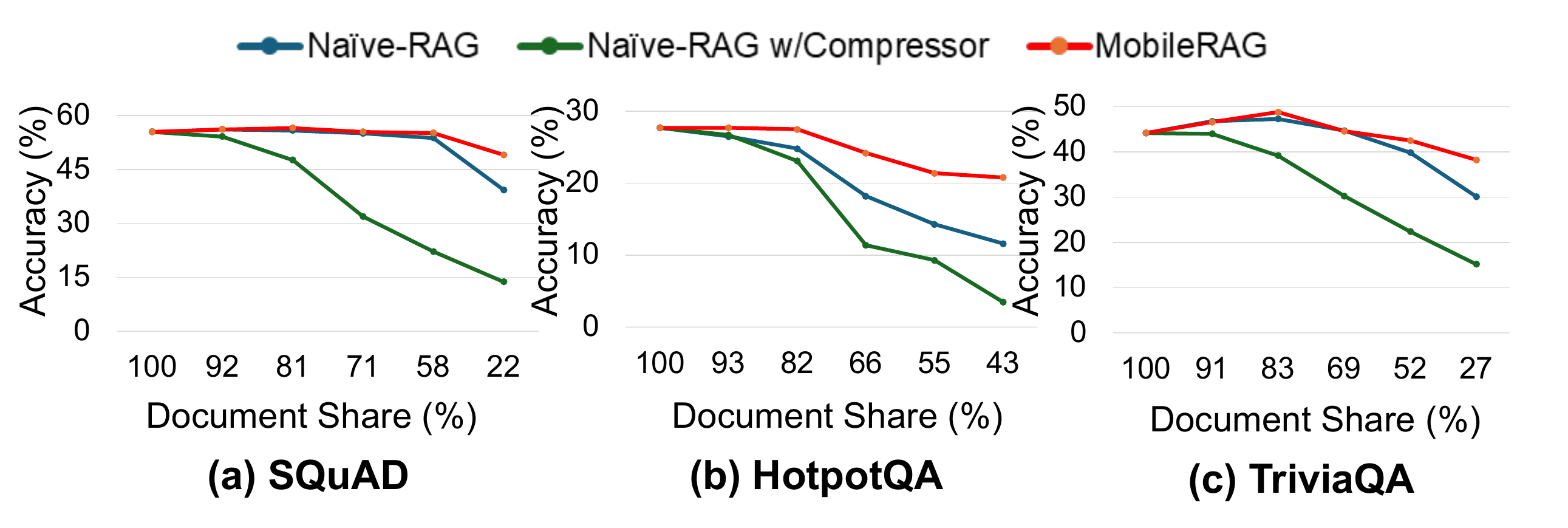}

  \caption{Comparison of SCR-based MobileRAG and Naive-RAG with Small Chunks or Compressor.}
\label{fig:SCR}
    \vspace*{-0.2cm}
\end{figure}

\subsubsection{Analysis for Accuracy:} Figure~\ref{fig:SCR} shows the performance of our SCR method across different sliding window sizes and overlap sizes on three datasets. The SCR method can reduce the input size for the SQuAD dataset by \texttt{42}\%, for HotpotQA by \texttt{7}\%, and for TriviaQA by \texttt{31\%} without any loss of accuracy. This is made possible by first retrieving $k$ relevant documents through the query–vector search stage, then applying the SCR method only to that already-filtered subset. In contrast, a compressor-based approach discards too much context and consequently suffers a steep drop in accuracy. Similarly, Naive-RAG with small chunk sizes from the outset approaches the SCR method's accuracy trend but ultimately lags behind, because preemptively shrinking chunks discards vital context and degrades RAG quality.

Additionally, Table~\ref{tab:scr_comparison} summarizes the average token count per document before and after applying the SCR method on each dataset. In these experiments, we set \textit{sliding\_window\_size} = 3, \textit{overlap\_size} = 2, and \textit{context\_extension\_size} = 1, resulting in windows of five sentences each ($\text{\textit{sliding\_window\_size}} + 2 \cdot \text{\textit{context\_extension\_size}}$). The results confirm that the SCR method achieves substantial reductions in token count while maintaining nearly the same accuracy, underscoring its effectiveness in lightweighting the retrieval subset. We adopt these parameter values for all subsequent experiments.

In Table~\ref{tab:MobileRAG}, we further compare the accuracy of Naive-RAG, Advanced RAG, EdgeRAG, and MobileRAG across the SQuAD, HotpotQA, and TriviaQA datasets. Although MobileRAG does not employ an explicit Re-Ranker model, it incorporates a reordering step at the end of the SCR method, which helps refine the document sequence. As a result, MobileRAG achieves higher accuracy than the non-reordering baselines (Naive-RAG and EdgeRAG) on all three datasets. Moreover, when compared to Advanced RAG, MobileRAG exhibits comparable accuracy, demonstrating that an efficient reordering strategy can significantly boost performance even without an additional large Re-Ranker model.

\begin{table}[t]
  \vspace*{-0.3cm}
\caption{Context token comparison pre/post-SCR.}
  \vspace*{-0.2cm}
\label{tab:scr_comparison}
\begin{tabular}{l|c|c|c}
\toprule
 & \textbf{SQuAD} & \textbf{HotpotQA} & \textbf{TriviaQA} \\
\midrule
\textbf{Before SCR} & 155 & 309 & 287 \\
\textbf{After SCR}  & 90 (-42\%) & 287 (-7\%) & 198 (-31\%) \\
\bottomrule
\end{tabular}
  \vspace*{-0.3cm}
\end{table}

\subsubsection{Analysis for Memory Usage:} \label{subsec:Analysis_Memory}
\begin{figure}[h]
    \vspace*{-0.3cm}
  \includegraphics[width=\linewidth]{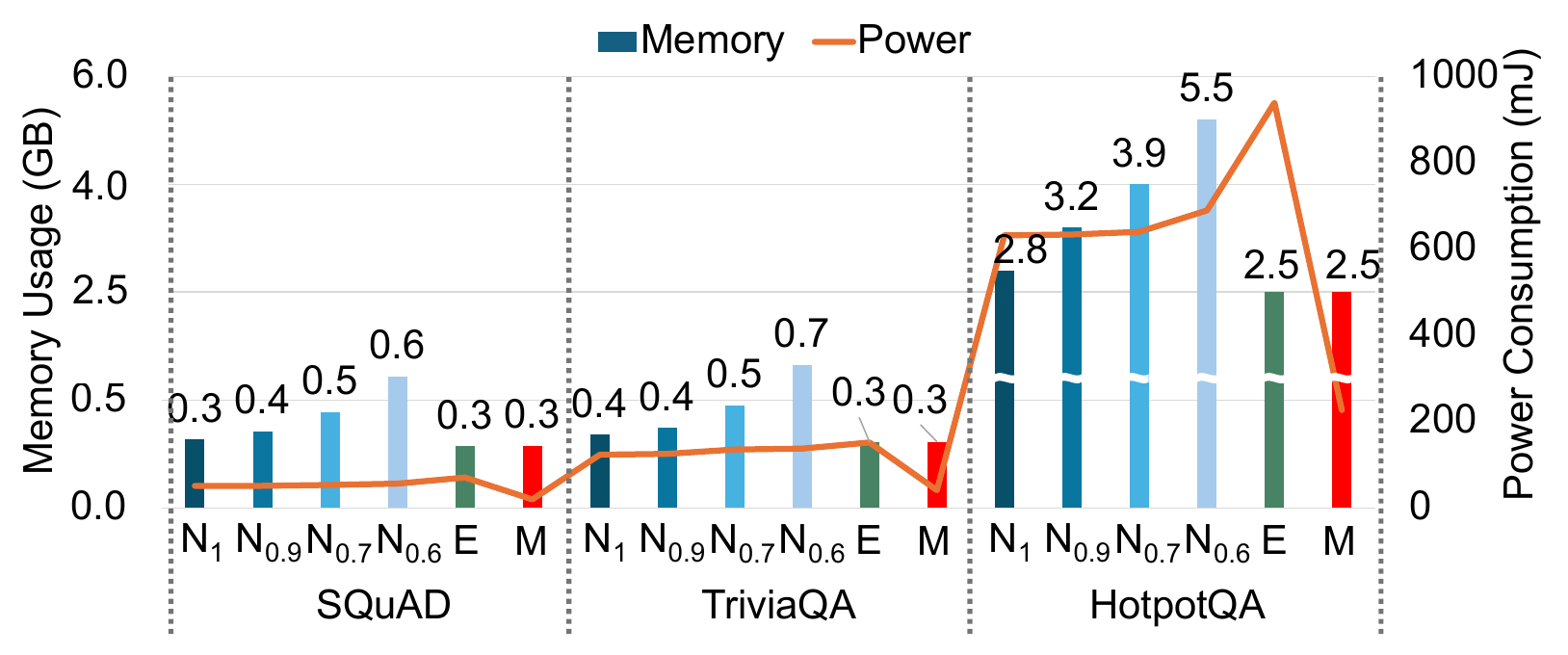}
    \caption{Memory and Power for retrieval: $N_{1}$, $N_{0.9}$, $N_{0.7}$, $N_{0.6}$ denote Naive-RAG at chunk ratios 1, 0.9, 0.8, 0.6; $E$ denotes EdgeRAG; $M$ denotes MobileRAG.}
  \label{fig:rag_comparison}
    \vspace*{-0.3cm}
\end{figure}

As shown in Figure~\ref{fig:SCR}, reducing the initial chunk size of the source documents can maintain a certain level of accuracy when fed into the sLM. However, using this approach for RAG i.e., naively shrinking all document chunks is detrimental to memory usage. According to Figure~\ref{fig:rag_comparison}, when the initial chunk size is reduced from the outset (\textit{e.g., $N_{0.6}$}) the memory consumption increases by 2.1X on SQuAD, 2.13X on TriviaQA, and 2.2X on HotpotQA compared to MobileRAG. Although EdgeRAG (which leverages IVF-DISK) exhibits slightly lower memory usage than MobileRAG, the difference is marginal. Overall, these results show that naive pre-retrieval chunk shrinking inflates memory usage, whereas targeted post-retrieval reduction strategies like the SCR method achieve better trade-offs between accuracy and resource efficiency.

\subsubsection{Analysis for sLM Inference Latency:}
Table~\ref{tab:MobileRAG} shows the TTFT, including sLM inference. By applying the SCR method to reduce input tokens, MobileRAG significantly reduces TTFT across all models and datasets. MobileRAG outperforms Naive-RAG and EdgeRAG by 10.4–26.2\%, and AdvancedRAG by 12.9–30.1\% in Qwen-2.5 0.5B; Naive-RAG and EdgeRAG by 14.9–35\%, and AdvancedRAG by 16.3–37.1\% in Qwen-2.5 1.5B; and Naive-RAG and EdgeRAG by 17.7–40.5\%, and AdvancedRAG by 18.5–41.6\% in Deepseek-r1 1.5B. These improvements are consistent across datasets, demonstrating MobileRAG's efficiency without sacrificing accuracy.

\begin{figure}[h]
    \vspace*{-0.3cm}
  \includegraphics[width=\linewidth]{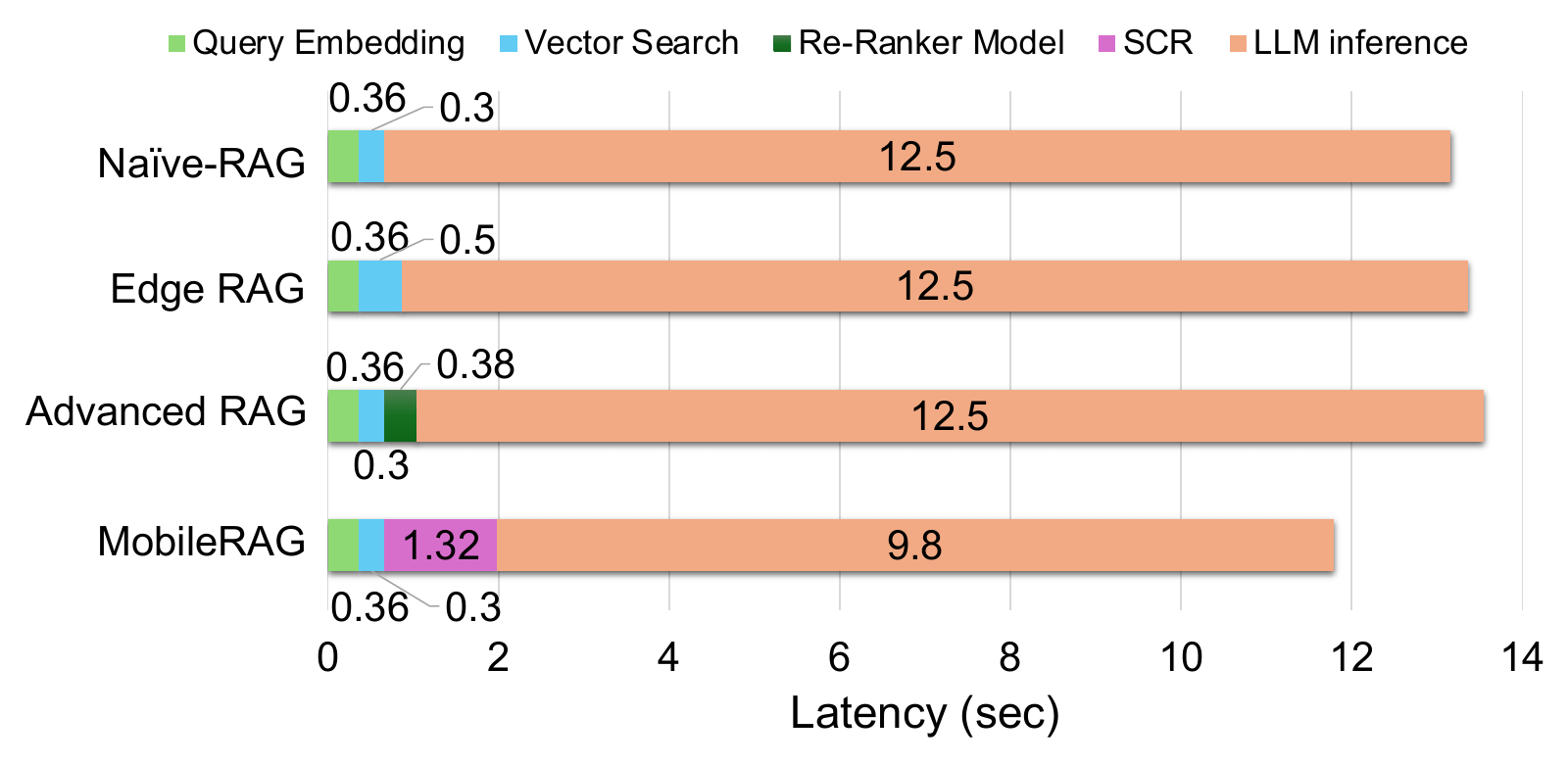}
  \caption{Breakdown of TTFT on HotpotQA Dataset.}
  \label{fig:TTFT}
\vspace*{-0.4cm}
\end{figure}

Figure~\ref{fig:TTFT} provides additional details on TTFT; while the SCR method itself adds a small overhead, the resultant decrease in sLM inference time ultimately reduces the total TTFT.

\subsubsection{Analysis for Power Consumption:} 
We divide our power consumption analysis into two parts: the retrieval stage and the full RAG pipeline. First, Figure~\ref{fig:rag_comparison} reports power consumption during the retrieval phase alone. Because EdgeRAG relies on IVF-based disk search, its slower query times lead to higher overall energy usage. In contrast, MobileRAG reduces power consumption by 61\%, 75\%, and 73\% on SQuAD, HotpotQA, and TriviaQA, respectively, compared to EdgeRAG. Likewise, compared to Naive-RAG, MobileRAG lowers power by 47\%, 64\%, 67\% on SQuAD, HotpotQA, and TriviaQA. Notably, as discussed in section~\ref{subsec:Analysis_Memory}, MobileRAG maintains similar memory usage compared to EdgeRAG yet achieves the lowest overall power consumption, demonstrating its effectiveness in resource-constrained mobile environments.

\begin{table*}[t]
\centering
\caption{RAG comparison up to generation on Accuracy(\%), TTFT(sec), and Power(J) per query.}
\vspace*{-0.2cm}
\label{tab:MobileRAG}
\begin{tabular}{l|l|c|cc|c|cc|c|cc}
\toprule
Model & Method & \multicolumn{3}{c|}{SQuAD} & \multicolumn{3}{c|}{HotpotQA} & \multicolumn{3}{c}{TriviaQA} \\
\cmidrule(lr){3-5}\cmidrule(lr){6-8}\cmidrule(lr){9-11}
 & & Acc & TTFT & Power & Acc & TTFT & Power & Acc & TTFT & Power \\
\midrule
{Qwen-2.5 0.5B} 
& Naive-RAG
  & 55.4 & 6.79 & 32.72
  & 27.7 & 13.16 & 61.40
  & 44.2 & 13.72 & 63.93 \\
& EdgeRAG
  & 55.4 & 6.79 & 32.75
  & 27.7 & 13.36 & 62.30
  & 44.2 & 13.73 & 65.40 \\
& Advanced RAG
  & 56.6 & 7.17 & 34.43
  & 27.9 & 13.54 & 63.11
  & 48.8 & 14.10 & 67.09 \\  
& MobileRAG
  & 56.5 & \textbf{5.01} & \textbf{24.71}
  & 27.7 & \textbf{11.79} & \textbf{55.21}
  & 48.8 & \textbf{10.59} & \textbf{51.30} \\
\midrule
{Qwen-2.5 1.5B}
& Naive-RAG
  & 63.2 & 11.40 & 86.24
  & 32.8 & 22.09 & 166.39
  & 56.3 & 23.25 & 175.08 \\
& EdgeRAG
  & 63.2 & 11.41 & 86.27
  & 32.9 & 22.29 & 167.89
  & 56.3 & 23.25 & 175.12 \\
& Advanced RAG
  & 65.1 & 11.78 & 89.09
  & 33.4 & 22.47 & 169.24
  & 56.9 & 23.63 & 177.93 \\
& MobileRAG
  & 65.1 & \textbf{7.41} & \textbf{56.28}
  & 33.2 & \textbf{18.79} & \textbf{141.65}
  & 56.7 & \textbf{16.94} & \textbf{127.76} \\
\midrule
{Deepseek-r1 1.5B}
& Naive-RAG
  & 63.2 & 19.71 & 107.08
  & 35.6 & 38.16 & 203.97
  & 55.1 & 40.39 & 215.67 \\
& EdgeRAG
  & 63.2 & 19.71 & 107.10
  & 35.6 & 38.36 & 205.02
  & 55.1 & 40.40 & 215.70 \\
& Advanced RAG
  & 65.1 & 20.09 & 109.07
  & 36.3 & 38.54 & 205.96
  & 55.8 & 40.77 & 217.67 \\
& MobileRAG
  & 65.5 & \textbf{11.73} & \textbf{65.18}
  & 36.0 & \textbf{31.40} & \textbf{168.46}
  & 55.8 & \textbf{28.36} & \textbf{152.52} \\
\bottomrule
\end{tabular}
\vspace*{-0.2cm}
\end{table*}

Table~\ref{tab:MobileRAG} goes beyond the retrieval phase and summarizes the total power consumption across the entire RAG pipeline, from retrieval to answer generation. The results clearly indicate that MobileRAG consistently achieves the lowest overall power consumption across different models and datasets. Specifically, it reduces power usage by up to 24.5\% in Qwen-2.5 0.5B, 34.7\% in Qwen-2.5 1.5B, and 39.1\% in Deepseek-r1 1.5B compared to Naive-RAG and EdgeRAG, and up to 28.2\%, 36.8\%, and 40.2\%, respectively, compared to AdvancedRAG. Notably, MobileRAG's efficiency advantage grows with model size, significantly curbing overall power usage.

Finally, Table~\ref{tab:battery} illustrates the real-world battery impact on a Galaxy~S24 device when running Qwen2.5~0.5B, Qwen2.5 1.5B, or Deepseek-r1~1.5B. The measured prompt evaluation speeds are approximately {90}, {50}, and {35} tokens/s, respectively, while the corresponding generation speeds reach {14.5}, {10}, and {9} tokens/s. Across these models, each {1k} tokens of processing consumes about {0.1\%}, {0.3\%}, and {0.36\%} of the phone’s battery, confirming that faster inference has a direct correlation with lower battery draw.

\begin{table}[h]
\vspace*{-0.2cm}
\caption{Prompt Evaluation Speed, Generation Speed and Battery Impact on Mobile Device.}
\vspace*{-0.2cm}
\label{tab:battery}
\resizebox{\columnwidth}{!}{%
\begin{tabular}{l|c|c|c}
\toprule
\multirow{1}{*}{\textbf{sLM}} 
& \shortstack{\textbf{Prompt}\\\textbf{Eval. Speed}} 
& \shortstack{\textbf{Generation}\\\textbf{Speed}} 
& \shortstack{\textbf{Battery}\\\textbf{Impact}} \\
& \textbf{(token/s)} 
& \textbf{(token/s)} 
& \textbf{(\% /1k tokens)} \\
\midrule
Qwen2.5 0.5B & 90 & 14.5 & 0.10 \\
Qwen2.5 1.5B & 50 & 10   & 0.30 \\
Deepseek 1.5B & 35 & 9    & 0.36 \\
\bottomrule
\end{tabular}
} 
\vspace*{-0.2cm}
\end{table}

\section{Related Work}
\label{sec:RelatedWork}

\begin{itemize}[labelindent=0em,labelsep=0.3em,leftmargin=*,itemsep=0em]

\item \textbf{Clustering Vector Search Methods:}
IVF clusters vectors (e.g., via k-means), limiting searches to relevant clusters for efficiency~\cite{nister2006scalable,babenko2014inverted}. However, large centroid and inverted list storage pose memory issues on mobile devices, often mitigated by Product Quantization (PQ) at increased computational cost~\cite{5432202,jegou2010product,McGowan2021,ge2013optimized,zhang2014composite,chen2010approximate}.

\item \textbf{Graph Vector Search Methods:}
HNSW employs multi-layered graphs for fast approximate neighbor searches using greedy traversal~\cite{8594636,liu2022optimizing,malkov2014approximate,malkov2018efficient,fu2016efanna,dong2011efficient,wang2021comprehensive,prokhorenkova2020graph}. Despite effectiveness, extensive graph structures require significant RAM, complicating mobile deployment. PQ compression reduces memory but introduces additional preprocessing overhead~\cite{5432202,jegou2010product,McGowan2021}.

\item \textbf{Naive-RAG:}
Directly retrieves documents via vector searc-h, feeding them straight into a Language Model \cite{NEURIPS2020_6b493230,guu2020retrieval,fan2025minirag,karpukhin2020dense,lee2019latent,khandelwal2019generalization,petroni2020kilt}. This approach is efficient yet suboptimal for ambiguous or specialized queries due to its single-pass nature\cite{hwang2024dslr,yu2025rankrag,gao2023retrieval}.

\item \textbf{Advanced RAG:}
Improves upon Naive-RAG by introducing a Re-Ranker to refine document relevance post-retrieval, significantly boosting output quality \cite{hwang2024dslr,yu2025rankrag,paranjape2021hindsight,shi2023replug,wang2023self,trivedi2022interleaving,glass2022re2g}. However, this additional refinement raises computational demand and latency.

\item \textbf{On-Device RAG:}
EdgeRAG optimizes mobile resource usage by combining IVF-DISK indexing and embedding caching, loading embeddings from disk as needed. Despite reduced RAM usage, feeding entire retrieved documents (e.g., \textit{2K} tokens) to the LM causes higher latency and power consumption, challenging mobile performance consistency~\cite{seemakhupt2024edgerag}.

\end{itemize}

\section{Conclusions} \label{sec:Conclusion}
In this paper, we proposed a fast, memory-efficient, and power-efficient method for on-device RAG by  combining the EcoVector index and the SCR method.
On actual mobile devices, it significantly outperforms existing methods, improving search latency by 1.72--8.89 times\,(at 0.93 recall@10 for SIFT), TTFT by 1.18--1.41 times, reducing memory consumption by 10.7--54.5\%, and decreasing power consumption by 24.4--40.2\%.
Future work will focus on exploring advanced NPU/GPU co-processing 
and various embedding strategies to enhance the viability of MobileRAG across diverse real-world scenarios.


\bibliographystyle{ACM-Reference-Format}
\bibliography{sigconf}

\appendix

\end{document}